\documentclass[]{IEEEtran}

\usepackage[numbers,sort&compress]{natbib}

\makeatletter

\usepackage{etex}
\usepackage{zref-savepos}

\newcounter{mnote}%[page]

\def\xmarginnote{%
  \xymarginnote{\hskip -\marginparsep \hskip -\marginparwidth}}

\def\ymarginnote{%
  \xymarginnote{\hskip\columnwidth \hskip\marginparsep}}

\long\def\xymarginnote#1#2{%
\vadjust{#1%
\smash{\hbox{{%
        \hsize\marginparwidth
        \@parboxrestore
        \@marginparreset
\footnotesize #2}}}}}

\def\mnoteson{%
\gdef\mnote##1{\refstepcounter{mnote}\label{##1}%
  \zsavepos{##1}%
  \ifnum20432158>\number\zposx{##1}%
  \xmarginnote{{\color{blue}\bf $\langle$\arabic{mnote}$\rangle$}}% 
  \else
  \ymarginnote{{\color{blue}\bf $\langle$\arabic{mnote}$\rangle$}}%
  \fi%
}
  }
\gdef\mnotesoff{\gdef\mnote##1{}}
\mnoteson
\mnotesoff

\newcommand{\figref}[1]{Fig.~\ref{#1}}

\usepackage{framed}
\usepackage{subcaption}
\usepackage{comment}
\usepackage[svgnames,dvipsnames]{xcolor}
\usepackage[all]{xy}
\usepackage{tikz}
\usetikzlibrary{positioning,matrix,through,calc,arrows,fit,shapes,decorations.pathreplacing,decorations.markings,}
\tikzstyle{block} = [draw,fill=blue!20,minimum size=2em]

\usepackage{qsymbols,amssymb,mathrsfs}
\usepackage[standard,thmmarks]{ntheorem}
\theoremstyle{plain}
\theoremsymbol{\ensuremath{_\vartriangleleft}}
\theorembodyfont{\itshape}
\theoremheaderfont{\normalfont\bfseries}
\theoremseparator{}

\theoremstyle{nonumberplain}
\theoremheaderfont{\scshape}
\theorembodyfont{\normalfont}
\theoremsymbol{\ensuremath{_\blacktriangleleft}}

\theoremnumbering{arabic}
\theoremstyle{plain}
\usepackage{latexsym}
\theoremsymbol{\ensuremath{_\Box}}
\theorembodyfont{\itshape}
\theoremheaderfont{\normalfont\bfseries}
\theoremseparator{}

\theorembodyfont{\upshape}
\theoremprework{\bigskip\hrule}
\theorempostwork{\hrule\bigskip}
\usepackage[overload]{empheq} % no \intertext and \displaybreak

\let\iftwocolumn\if@twocolumn
\g@addto@macro\@twocolumntrue{\let\iftwocolumn\if@twocolumn}
\g@addto@macro\@twocolumnfalse{\let\iftwocolumn\if@twocolumn}

\mathtoolsset{showonlyrefs=false,showmanualtags}
\let\underbrace\LaTeXunderbrace % adapt spacing to font sizes

\renewcommand{\eqref}[1]{\textup{(\refeq{#1})}} % eqref was not allowed in
\newtagform{brackets}[\textbf]{(}{)}   % new tags for equations
\usetagform{brackets}
\usepackage[Smaller]{cancel}

\PassOptionsToPackage{breaklinks,letterpaper,hyperindex=true,backref=false,bookmarksnumbered,bookmarksopen,linktocpage,colorlinks,linkcolor=BrickRed,citecolor=OliveGreen,urlcolor=Blue,pdfstartview=FitH}{hyperref}
\usepackage{hyperref}

\usepackage{graphicx,psfrag}
\graphicspath{{figure/}{image/}} % Search path of figures

\usepackage{diagbox} % \backslashbox{}{} for slashed entries
\usepackage{algorithm2e}
\usepackage{listings} 
\lstdefinelanguage{Maple}{
  morekeywords={proc,module,end, for,from,to,by,while,in,do,od
    ,if,elif,else,then,fi ,use,try,catch,finally}, sensitive,
  morecomment=[l]\#,
  morestring=[b]",morestring=[b]`}[keywords,comments,strings]
\DeclareMathAlphabet{\mathpzc}{OT1}{pzc}{m}{it}
\usepackage{upgreek} % \upalpha,\upbeta, ...
\usepackage{dsfont}  % \mathds

\def\multi@nostar#1#2{%
  \expandafter\def\csname multi#1\endcsname##1{%
    \if ##1.\let\next=\relax \else
    \def\next{\csname multi#1\endcsname}     
    \expandafter\newcommand\csname #1##1\endcsname{#2}
    \fi\next}}

\def\multi@star#1#2{%
  \expandafter\def\csname #1\endcsname##1{#2}
  \multi@nostar{#1}{#2}
}

\newcommand{\multi}{%
  \@ifstar \multi@star \multi@nostar}

\multi*{rm}{\mathrm{#1}}
\multi*{mc}{\mathcal{#1}}
\multi*{op}{\mathop {\operator@font #1}}
\multi*{ds}{\mathds{#1}}
\multi*{set}{\mathcal{#1}}
\multi*{rsfs}{\mathscr{#1}}
\multi*{pz}{\mathpzc{#1}}
\multi*{M}{\boldsymbol{#1}}
\multi*{R}{\mathsf{#1}}
\multi*{RM}{\M{\R{#1}}}
\multi*{bb}{\mathbb{#1}}
\multi*{td}{\tilde{#1}}
\multi*{tR}{\tilde{\mathsf{#1}}}
\multi*{trM}{\tilde{\M{\R{#1}}}}
\multi*{tset}{\tilde{\mathcal{#1}}}
\multi*{tM}{\tilde{\M{#1}}}
\multi*{baM}{\bar{\M{#1}}}
\multi*{ol}{\overline{#1}}

\multirm  ABCDEFGHIJKLMNOPQRSTUVWXYZabcdefghijklmnopqrstuvwxyz.
\multiol  ABCDEFGHIJKLMNOPQRSTUVWXYZabcdefghijklmnopqrstuvwxyz.
\multitR   ABCDEFGHIJKLMNOPQRSTUVWXYZabcdefghijklmnopqrstuvwxyz.
\multitd   ABCDEFGHIJKLMNOPQRSTUVWXYZabcdefghijklmnopqrstuvwxyz.
\multitset ABCDEFGHIJKLMNOPQRSTUVWXYZabcdefghijklmnopqrstuvwxyz.
\multitM   ABCDEFGHIJKLMNOPQRSTUVWXYZabcdefghijklmnopqrstuvwxyz.
\multibaM   ABCDEFGHIJKLMNOPQRSTUVWXYZabcdefghijklmnopqrstuvwxyz.
\multitrM   ABCDEFGHIJKLMNOPQRSTUVWXYZabcdefghijklmnopqrstuvwxyz.
\multimc   ABCDEFGHIJKLMNOPQRSTUVWXYZabcdefghijklmnopqrstuvwxyz.
\multiop   ABCDEFGHIJKLMNOPQRSTUVWXYZabcdefghijklmnopqrstuvwxyz.
\multids   ABCDEFGHIJKLMNOPQRSTUVWXYZabcdefghijklmnopqrstuvwxyz.
\multiset  ABCDEFGHIJKLMNOPQRSTUVWXYZabcdefghijklmnopqrstuvwxyz.
\multirsfs ABCDEFGHIJKLMNOPQRSTUVWXYZabcdefghijklmnopqrstuvwxyz.
\multipz   ABCDEFGHIJKLMNOPQRSTUVWXYZabcdefghijklmnopqrstuvwxyz.
\multiM    ABCDEFGHIJKLMNOPQRSTUVWXYZabcdefghijklmnopqrstuvwxyz.
\multiR    ABCDEFGHIJKL NO QR TUVWXYZabcd fghijklmnopqrstuvwxyz.
\multibb   ABCDEFGHIJKLMNOPQRSTUVWXYZabcdefghijklmnopqrstuvwxyz.
\multiRM   ABCDEFGHIJKLMNOPQRSTUVWXYZabcdefghijklmnopqrstuvwxyz.

\newcommand{\dotleq}{\buildrel \textstyle  .\over {\smash{\lower
      .2ex\hbox{\ensuremath\leqslant}}\vphantom{=}}}
\newcommand{\dotgeq}{\buildrel \textstyle  .\over {\smash{\lower
      .2ex\hbox{\ensuremath\geqslant}}\vphantom{=}}}

\newcommand{\bM}{\begin{bmatrix}}
\newcommand{\eM}{\end{bmatrix}}
\newcommand{\bSM}{\left[\begin{smallmatrix}}
\newcommand{\eSM}{\end{smallmatrix}\right]}
\renewcommand*\env@matrix[1][*\c@MaxMatrixCols c]{%
  \hskip -\arraycolsep
  \let\@ifnextchar\new@ifnextchar
  \array{#1}}

\newqsymbol{`N}{\mathbb{N}}
\newqsymbol{`R}{\mathbb{R}}
\newqsymbol{`P}{\mathbb{P}}
\newqsymbol{`Z}{\mathbb{Z}}

\newqsymbol{`|}{\mid}
\newqsymbol{`8}{\infty}
\newqsymbol{`1}{\left}
\newqsymbol{`2}{\right}
\newqsymbol{`6}{\partial}
\newqsymbol{`0}{\emptyset}
\newqsymbol{`-}{\leftrightarrow}
\newqsymbol{`<}{\langle}
\newqsymbol{`>}{\rangle}

\DeclarePairedDelimiter\abs{\lvert}{\rvert}

\DeclarePairedDelimiter\Set{\{}{\}}
\newcommand{\imod}[1]{\allowbreak\mkern10mu({\operator@font mod}\,\,#1)}

\newcommand{\threecols}[3]{
\hbox to \textwidth{%
      \normalfont\rlap{\parbox[b]{\textwidth}{\raggedright#1\strut}}%
        \hss\parbox[b]{\textwidth}{\centering#2\strut}\hss
        \llap{\parbox[b]{\textwidth}{\raggedleft#3\strut}}%
    }% hbox 
}

\newcommand{\reason}[2][\relax]{
  \ifthenelse{\equal{#1}{\relax}}{
    \left(\text{#2}\right)
  }{
    \left(\parbox{#1}{\raggedright #2}\right)
  }
}

\newcommand{\utag}[2]{\mathop{#2}\limits^{\text{(#1)}}}
\newcommand{\uref}[1]{(#1)}
\makeatother

\usepackage{ctable}
\usepackage{fouridx}
\usepackage{framed}
\usetikzlibrary{positioning,matrix}

\usepackage{paralist}
\usepackage{enumerate}

\usepackage[normalem]{ulem}

\numberwithin{equation}{section}
\makeatletter
\@addtoreset{equation}{section}
\renewcommand{\theequation}{\arabic{section}.\arabic{equation}}
\@addtoreset{Theorem}{section}
\renewcommand{\theTheorem}{\arabic{section}.\arabic{Theorem}}
\@addtoreset{Lemma}{section}
\renewcommand{\theLemma}{\arabic{section}.\arabic{Lemma}}
\@addtoreset{Corollary}{section}
\renewcommand{\theCorollary}{\arabic{section}.\arabic{Corollary}}
\@addtoreset{Example}{section}
\renewcommand{\theExample}{\arabic{section}.\arabic{Example}}
\@addtoreset{Remark}{section}

\@addtoreset{Proposition}{section}
\renewcommand{\theProposition}{\arabic{section}.\arabic{Proposition}}
\@addtoreset{Definition}{section}

\@addtoreset{Claim}{section}

\@addtoreset{Subclaim}{Theorem}

\makeatother

\usepackage[inline]{enumitem}

\newcommand{\RS}{R_{\op{S}}}
\newcommand{\rK}{r_{\op{K}}}
\newcommand{\CS}{C_{\op{S}}}
\title{Secret Key Generation for\\ Minimally Connected Hypergraphical Sources}
\author{Qiaoqiao Zhou and Chung Chan 
        \thanks{Parts of this work were presented at the 2018 IEEE International Symposium on Information Theory (ISIT 2018), Vail, Colorado, U.S.A.~\cite{isit18}.}
        \thanks{Q.\ Zhou is with the Institute of Network Coding and the Department of Information Engineering, The Chinese University of Hong Kong, Hong Kong SAR, China (email: zq115@ie.cuhk.edu.hk).}	
	\thanks{C.\ Chan is with the Department of Computer Science, City University of Hong Kong, Hong Kong SAR, China (email: chung.chan@cityu.edu.hk).}
}
\begin{document}
\IEEEoverridecommandlockouts
\maketitle
\begin{abstract}
This paper investigates the secret key generation in the multiterminal source model, where users observing correlated sources discuss interactively under limited rates to agree on a secret key. We focus on a class of sources representable by minimally connected hypergraphs. For such sources, we give a single-letter explicit characterization of the region of achievable secret key rate and public discussion rate tuple. This is the first result that completely characterizes the achievable rate region for a multiterminal source model, which is beyond the PIN model on a tree. We also obtain an explicit formula for the maximum achievable secret key rate, called the constrained secrecy capacity, as a function of the total discussion rate. %The region is not only single-letter and the formula can be computed explicitly in polynomial-time.%Their goal is to agree upon a secret key that independent of the public discussion 
%As by-product of our analyses, various combinatorial properties of minimally connected hypergraph are established.
\end{abstract}
\section{Introduction}
\label{sec:introduction}
We consider the secret key generation problem among multiple users~\cite{csiszar04}, in which each user observes a distinct component of a correlated discrete memoryless multiple source. The users are allowed to discuss over a noiseless public channel, possibly interactively in several rounds, in order to agree upon a common secret key that is independent of their discussion. While the maximum achievable secret key rate with unlimited discussion rate was characterized in~\cite{csiszar04}, it remains open when the discussion has limited rate. 

The secret key generation problem under limited discussion rate was first studied by Csisz\'ar and Narayan for discrete sources in the two-user case with a helper~\cite{csiszar00}. For the one-way discussion, they characterized the optimal trade-off between the secret key rate and discussion rate. Their result was subsequently extended to Gaussian sources in~\cite{watanabe10,watanabe11}. The minimum overall rate of interactive discussion required to generate a secret key of maximum rate, called the \emph{communication complexity}, was examined by~\cite{tyagi13} in the two-user case, where they obtained a multi-letter characterization.~\cite{MKS16} extended their framework to the multiterminal case and obtained a multi-letter lower bound on the communication complexity. The bound was single-letterized recently by~\cite{chan17oo}. In~\cite{chan17isit,chan18LB}, the optimal trade-off between the secret key rate and total discussion rate was completely characterized for the pairwise independent network (PIN) model proposed in~\cite{nitinawarat-ye10, nitinawarat10}. In~\cite{courtade16}, a hypergraphical source model~\cite{chan10md} was considered, and each user observes one realization of the source. They determined the minimum amount of total discussion needed to generate a secret key of a given size when the discussion is restricted to be linear function of the source. However, their expression is NP-hard to compute. Determining each individual public discussion rate, namely the public discussion rate region, required to generate a given secret key rate was addressed by~\cite{LCV16} in the two-user case. They obtained a multi-letter characterization of the region of achievable secret key rate and public discussion rate tuple. In~\cite{liu15}, the achievable rate region was characterized for a variant of the multiterminal source model. In~\cite{chan17isit}, an outer bound on the achievable rate region was established for general multiterminal source model. The bound was shown to be tight for the PIN model on a tree, but remains unknown whether it is tight for other sources. Besides, although the expression is single-letter, it may take doubly exponential time to compute the tightest bound. %since you have to evaluate all the subsets and all partitions of each subset. 

In this paper, we study the public discussion rate region required to generate a given secret key rate in the multiterminal case. At the outset we must mention that such a characterization, even for the two-terminal case appears intractable~\cite{LCV16}. On the other hand, in the multiterminal case, even the communication complexity of the hypergraphical source model remains unknown~\cite{MKS16,mukherjee16,chan17isit,chan17oo,chan18LB}. Therefore, we shall focus on a class of multiterminal sources that can be represented by minimally connected hypergraphs. For such kind of sources, we give a \emph{single-letter} explicit characterization of the region of achievable secret key rate and public discussion rate tuple. %Since MCH contains tree as a special case, our result generalizes~\cite{chan17isit}. 
      We would like to highlight that this is the first result which completely characterizes the achievable rate region for a large class of multiterminal sources, beyond the PIN model on a tree, without any restriction on the number of rounds of interactive discussion. %Also, the region does not involve any auxiliary random variables.
      Besides, we also obtain an explicit formula for the maximum achievable secret key rate under any given total discussion rate, referred to as the constrained secrecy capacity. Towards deriving the main results, we clarify some combinatorial properties of hypergraphs. More precisely, we find a way to reduce a minimally connected hypergraph  to a hypertree. This reduction guarantees the existence of unique paths among certain sets of vertices, which, in turn, gives rise to our capacity-achieving scheme that propagates the secret key along those paths, similar to the tree-packing protocol for the PIN model in~\cite{nitinawarat-ye10,nitinawarat10}. Furthermore, %to identify the extreme points, 
%we show that the number of connected components of certain subhypergraph is supermodular. 
we show that for certain subhypergraphs, the number of connected components is supermodular, while for some other parts, the normalization of the number of connected components by minus one is subadditive. With these, we get a characterization of the achievable rate region, which is specified concisely by a minimal set of inequalities. The combinatorial structures we show are general and fundamental, and may be of independent interest from a graph-theoretic viewpoint.
 
The paper is organized as follows. Section~\ref{sec:problem} formulates the problem and defines the hypergraphical source model. In Section~\ref{sec:preliminaries},  we further introduce some preliminaries on hypergraphs. The main results of the paper are presented in Section~\ref{sec:results} and proved in Section~\ref{sec:results:proof}. Finally, in Section~\ref{sec:con}, we conclude the paper with some remarks. %The proofs of some of the technical results appear in the appendices.

\section{Problem Formulation}
\label{sec:problem}

Let $V:=[\abs{V}]:=\{1,2,\ldots,\abs{V}\}$ be a finite set of $\abs{V}\geq2$ users. The users have access to a correlated discrete memoryless multiple source $$\RZ_V:= (\RZ_i\mid i\in V)$$ taking values from a finite set $$Z_V:= \prod\nolimits_{i\in V} Z_i.$$ We use the sans serif font for random variable and the usual math italic font for its corresponding alphabet set. The users try to generate a secret key via public discussion as follows. First,
each user $i\in V$ observes a sequence of $n$ i.i.d. samples $$\RZ_i^n:=(\RZ_{it}\mid t\in[n])$$ of his source $\RZ_i$.
Then, each user $i\in V$ generates a private randomization variable $\RU_i$ that is independent of all other randomness, i.e.,
	\begin{align*}
		P_{\RU_V|\RZ_V^n}=\prod\nolimits_{i\in V} P_{\RU_i}.
	\end{align*}
Following these observations, the users are allowed to discuss interactively over a noiseless public channel. We assume without any loss of generality that the users take turn to discuss for $\eta$ number of rounds.\footnote{Here, $\eta$ does not depend on $n$ and can be any finite positive integer.} More specifically, at round $t\in[\eta]$, each user $i\in V$ reveals a message that is a function of its accumulated observations, namely, 
%\begin{subequations}
\begin{align*}
	\label{eq:discussion}
\RF_{it}:=f_{it}(\RU_i,\RZ_i^n,\RF_{[i-1]t},\RF_V^{t-1}),
\end{align*}
where $$\RF_{[i-1]t}:=(\RF_{jt}\mid j<i)$$ denotes all the previous messages in the same round, and $$\RF_V^{t-1}:=(\RF_{i \tau }\mid i\in V, \tau<t)$$ denotes all the messages in the previous rounds. We will write
\begin{align*}
\RF_i :=(\RF_{it}\mid t\in[\eta]) \quad\text{and}\quad \RF:= (\RF_i\mid i\in V)\label{eq:F}
\end{align*}
%\end{subequations}
to denote, respectively, the collection of messages from user $i\in V$ and all users. The discussion is said to be \emph{linear} if all functions $f_{it}, i\in V, t\in[\eta]$ are linear, and \emph{non-interactive} if $\RF_i=f_i(\RU_i,\RZ_i^n), i\in V$ instead. 
After the public discussion, each user $i\in V$ then try to extract a common secret key $\RK$ from its accumulated observations. The secret key is required to satisfy
%\begin{subequations}
%\label{eq:weak}
\begin{align}
\text{Pr}(\exists i\in V, \RK\neq\psi_i(\RU_i,\RZ_i^n,\RF))\leq\epsilon_n, \label{eq:recover}\\
\frac{1}{n}[\log\abs{K}-H(\RK|\RF)]\leq\delta_n, \label{eq:secrecy}
\end{align}
%\end{subequations}
for some function $\psi_i$ for each $i\in V$ and for some $\epsilon_n, \delta_n\to 0$ as $n\to\infty$. The conditions~\eqref{eq:recover} and~\eqref{eq:secrecy} correspond to the concept of \emph{weak secrecy}. 
We say that a rate tuple $(\rK,r_V)$ where $r_V:=(r_i\mid i\in V)$ is \emph{achievable} if there exists a sequence of $(\RU_V,\RF,\RK)$ in $n$ satisfying 
	\begin{subequations}
	\begin{align}
	0\leq\rK\leq\liminf_{n\to\infty}\frac{1}{n}\log\abs{K}, \text{ and } 
	\end{align}
	\begin{align}
	r_i\geq\limsup_{n\to\infty}\frac{1}{n}\log\abs{F_i},\quad\forall i\in V. \label{eq:rate}
	\end{align}
	\end{subequations}
	in addition to \eqref{eq:recover} and \eqref{eq:secrecy}. Furthermore, $(\rK,r_V)$ or simply $\rK$ is said to be attained with \emph{zero error} and \emph{perfect secrecy} if for some $n$
\begin{align}
\label{eq:perfect}
\epsilon_n=\delta_n= 0.
\end{align}
	The \emph{achievable rate region} $\rsfsR$ is defined as
	\begin{align}
	\label{eq:def:region}
	\rsfsR:=\Set{(\rK,r_V)\in\mathbb{R}_+^{|V|+1}\mid (\rK,r_V)\kern0.5em \text{is achievable}}.
	\end{align}
	The maximum achievable secret key rate under a given total pubic discussion rate $R\geq 0$, referred to as the \emph{constrained secrecy capacity}, is then defined as 
	\begin{equation}
	\label{CS(R)}
	\CS(R):=\max\{\rK\mid (\rK,r_V)\in\rsfsR, r(V)\leq R\},
	\end{equation}
	where we write 
	\begin{align*}
r(B):=\sum_{i\in B} r_i \quad\text{for $B\subseteq V$}
  	\end{align*}
for notational convenience.
The \emph{unconstrained secrecy capacity} $\CS(\infty)$ defined and characterized in \cite{csiszar04} is
\begin{align}
\label{eq:cs8}
\CS(\infty):=\lim_{R\to\infty}\CS(R).
\end{align}
The \emph{communication complexity} $\RS$~\cite{MKS16} refers to the minimum total discussion rate required to achieve the unconstrained secrecy capacity, namely, 
\begin{align}
\label{eq:CS}
\RS:= \min\{R\geq 0\mid \CS(R)=\CS(\infty)\}.
\end{align}

If there is no restriction on the number of rounds of interactive discussion, single-letter characterizations of $\RS$ and $\CS(R)$ even for general two-terminal sources are not known~\cite{tyagi13,LCV16}, let alone such characterization of $\rsfsR$ for general multiterminal sources. To simplify the problem, the work in~\cite{chan17oo,MKS16,mukherjee16,chan17isit,courtade16,chan18LB} considered the hypergraphical source model in~\cite{chan10md}, which generalizes the PIN model in~\cite{nitinawarat-ye10,nitinawarat10}.  
\begin{Definition}[\mbox{\cite{chan10md}}]
\label{def:hs}
$\RZ_V$ is a hypergraphical source if there is a hypergraph $\mcH=(V,E,\xi)$ with an edge function $\xi: E\to2^V\setminus \{\emptyset\}$ and some mutually independent (hyper) edge random variables $\RX_{\rm{e}}$ for $\rm{e}\in E$, such that
\begin{align}
\label{eq:hs:Z_i}
\RZ_i:=(\RX_{\rm{e}}\mid \rm{e}\in E, i\in \xi(\rm{e})),\quad\forall i\in V.
\end{align}
The weight function $w: E\to\mathbb{R}_+$ of the hypergraph is then defined as
\begin{align}
\label{eq:hyper:weight}
w(\rm{e}):=H(\RX_{\rm{e}}), \quad \forall \rm{e}\in E.
\end{align}
%We assume $H(\RX_{\rm{e}})>0$ for all $\rm{e}\in E$ without loss of generality.
For simplicity, we assume that $\RX_{\rm{e}}$ is uniformly distributed over the set of all binary strings of length $w(\rm{e})>0$ (bits) for all $\rm{e}\in E$.
\end{Definition}

\begin{figure}
\centering
\tikzstyle{dot}=[shape=circle,dashed,draw=gray!100,thick,inner sep=1pt,minimum size=11pt]
\begin{tikzpicture}[]
	\node[dot] (4) at (0,0) {$4$};
	\node[dot] (2) at (0.95,0.2) {$2$};
	\node[dot] (5) at (2.0,0) {$5$};
	\node[dot] (6) at (0.95,-1.65) {$6$};
	\node[dot] (3) at (2.5,-1.2) {$3$};
 	\node[dot] (1) at (-0.6,-1.2) {$1$};
   	\draw [thick,red,rotate around={40:(0,-0.4)}](0,-0.44) ellipse (1.44cm and 0.62cm);
    	\draw [thick,blue,rotate around={140:(1.9,-0.4)}](1.9,-0.38) ellipse (1.44cm and 0.62cm);
    	\draw[thick,violet] (1,-1.3) ellipse (2cm and 0.6cm);
	\node[] () at (-.8,0)  {$\rm{a}$};
	\node[] () at (2.8,0) {$\rm{b}$};
	\node[] () at (0.95,-2.1)  {$\rm{c}$};
\end{tikzpicture}
\caption{A hypergraph $\mcH$ with $V=\Set{1,2,3,4,5,6}$, $E=\Set{\rm{a},\rm{b}, \rm{c}}$ and $\xi(\rm{a})=\Set{1,2,4}, \xi(\rm{b})=\Set{2,3,5}, \xi(\rm{c})=\Set{1,3,6}$. Dashed lines indicate the vertices while solid lines indicate the hyperedges.}\label{fig:mch}
\end{figure}
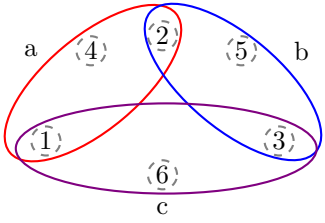

The following is an example of a hypergraphical source. 
\begin{Example}
\label{ex:mch}
Let $V=\Set{1,2,3,4,5,6}$ and define
\begin{equation*}
	\begin{aligned}
		\RZ_1&:=(\RX_{\rm{a}},\RX_{\rm{c}}),&   
		\RZ_2&:=(\RX_{\rm{a}},\RX_{\rm{b}}),&
		\RZ_3&:=(\RX_{\rm{b}},\RX_{\rm{c}}),\\
		\RZ_4&:=\RX_{\rm{a}},&
		\RZ_5&:=\RX_{\rm{b}},&
		\RZ_6&:=\RX_{\rm{c}}, 
	\end{aligned}
\end{equation*}
where $\RX_{i}$'s are independent with $H(\RX_{\rm{a}})=1, H(\RX_{\rm{b}})=3$, and $H(\RX_{\rm{c}})=2$. This is a hypergraphical source where the corresponding hypergraph is $\mcH$ in~\figref{fig:mch}. The weight function is $w(\rm{a})=H(\RX_{\rm{a}})=1$, $w(\rm{b})=H(\RX_{\rm{b}})=3$, and $w(\rm{c})=H(\RX_{\rm{c}})=2$. 
\end{Example}

For such hypergraphical sources, even the problem of characterizing $\RS$ remains open. Therefore, we further simplify the problem by restricting our attention to a hypergraphical source model defined below.
\begin{Definition}
A hypergraph $\mcH=(V,E,\xi)$ is said to be  \emph{connected} iff $\forall  C\subsetneq V$ with $C\neq \emptyset$, $\exists \rm{e}\in E$ such that $\xi(\rm{e})\cap C\neq \emptyset$ and $\xi(\rm{e})\setminus C\neq \emptyset$.
\end{Definition}
\begin{Definition}
\label{def:mch}
A connected hypergraph $\mcH=(V,E,\xi)$ is a \emph{minimally connected hypergraph (MCH)} iff it becomes disconnected after removing an arbitrary edge, i.e., $(V,E\setminus \{\rm{e}\},\xi)$ is disconnected for all $\rm{e}\in E$.
\end{Definition}
\begin{Definition}
\label{def:mchs}
$\RZ_V$ is a minimally connected hypergraphical (MCH) source if it is a hypergraphic source and the corresponding hypergraph is minimally connected.
\end{Definition}

The source in Example~\ref{ex:mch} is indeed a MCH source since removing edges $\rm{a},\rm{b}$, and $\rm{c}$ respectively disconnects nodes $4, 5$, and $6$ from other nodes.
Our goal is to characterize $\CS(R)$ and $\rsfsR$ for the above MCH source model.
\section{Preliminaries}
\label{sec:preliminaries}
In this section, we shall give a brief introduction of some hypergraph notions and operations that will be needed for the statements and proofs of our main results.

Let $\mcH=(V,E,\xi)$ be a hypergraph with a set of vertices $V=V(\mcH)$, a set of (hyper) edges $E=E(\mcH)$, and an edge function $\xi=\xi_{\mcH}: E\to2^V\setminus \{\emptyset\}$\footnote{We allow a hypergraph to have repeated edges, i.e., multiple edges covering the same set of vertices.}.
The \emph{degree} of a vertex $v\in V(\mcH)$ in $\mcH$, denoted by $\mathtt{d}_{\mcH}(v)$, is the number of incident edges associated with it, i.e.,
\begin{subequations}
\begin{align}
\label{eq:deg:v}
\mathtt{d}_{\mcH}(v):=\abs*{\Set{\rm{e}\in E(\mcH)\mid v\in \xi_{\mcH}(\rm{e})}}.
\end{align}
Similarly, for a set of vertices $C\subseteq V(\mcH)$, its degree is 
\begin{align}
\label{eq:deg:c}
\mathtt{d}_{\mcH}(C):=\abs{\Set{\rm{e}\in E(\mcH)\mid C\cap \xi_{\mcH}(\rm{e})}\neq`0}.
\end{align}
\end{subequations}
A \emph{loop} in $\mcH$ is an edge $\rm{e}\in E(\mcH)$ such that $|\xi_{\mcH}(\rm{e})|=1$.
A \emph{path} in $\mcH$ between two vertices $v_1$ and $v_\ell$ is a sequence $(v_1, \rm{e}_1, v_2,  \dots, \rm{e}_{\ell-1}, v_\ell)$ with the following properties: $\ell$ is a positive integer $\geq2$; $v_i\in V(\mcH)$ for $i=1,2,\dots, \ell$; all $v_i$ are distinct;  $\rm{e}_j\in E(\mcH)$ and $v_{j}, v_{j+1}\in \xi_{\mcH}(\rm{e}_j)$ for $j=1,2,\dots, \ell-1$; all $\rm{e}_j$ are distinct. The sequence is called a \emph{(Berge) cycle}~\cite{berge1989hypergraphs} if $v_1=v_\ell$ instead with $\ell\geq 3$. It follows from definition that a loop is not a cycle.\footnote{In graph theory literature, a loop is sometimes also considered as a cycle, i.e., the definition of a cycle requires $\ell\geq 2$ but not $\ell\geq 3$. We excluded loops in our definition to simplify the presentation of our main results and proofs.} We write $v_1\sim_{\mcH} v_\ell$ to indicate $v_\ell$ is reachable from $v_1$ via a path in $\mcH$. It is easy to see that $\sim_{\mcH}$ is an equivalence relation.  The equivalence relation $\sim_{\mcH}$ divides $V(\mcH)$ into a set of equivalence classes, each of which is the vertex set of a connected component of $\mcH$. Let $\kappa(\mcH)$ denote the number of equivalence classes, i.e., the number of connected components. A hypergraph $\mcH$ is connected iff %there is a path in $\mcH$ between any two distinct vertices, i.e., $v_i\sim_{\mcH} v_j, \forall v_i, v_j \in V(\mcH): v_i\neq v_j$; or equivalently $\kappa(\mcH)=1$.
$\kappa(\mcH)=1$. A special type of connected hypergraph, called hypertree, %that generalizes tree in graph, 
will be considered.
\begin{Definition}
\label{def:HT}
A hypergraph $\mcH$ is a hypertree iff $\mcH$ is connected, loopless, and the path between any pair of distinct vertices is unique. In other words, $\mcH$ contains neither loops nor cycles.
\end{Definition}

Our definition of hypertree appears new. It is a straightforward generalization of tree for hypergraph and is different from the standard definition~\cite{berge1989hypergraphs}.\footnote{%Our definition of hypertree is different from the standard definition. In hypergraph theory~\cite{berge1989hypergraphs}, %a hypergraph $\mcH$ is called a hypertree or arboreal hypergraph if it admits a host graph $\mcT$ such that $\mcT$ is a tree. In other words, 
In~\cite{berge1989hypergraphs}, a hypergraph $\mcH$ is a hypertree (also called arboreal hypergraph) iff there exists a tree $\mcT$ whose set of vertices is the same as that of $\mcH$ and such that every hyperedge of $\mcH$ induces a connected subtree in $\mcT$. Compared with the standard definition, our definition is more stringent, and therefore is a special case of the standard definition.}
An example of a hypertree is given in~\figref{fig:ht}.
\begin{figure}
\centering
	\tikzstyle{dot}=[shape=circle,dashed,draw=gray!100,thick,inner sep=1pt,minimum size=11pt]
\begin{tikzpicture}
	\node[dot] (1) at (-0.2,0) {$1$};
	\node[dot] [right= 0.8cm of 1] (2) {$2$};
	\node[dot] [right= 0.8cm of 2] (3) {$3$};
	\node[dot] [right= 0.8cm of 3] (4) {$4$};
	\node[dot] [right= -1.8cm of 1] (5) {$5$};
        \draw [thick,red](1.0,0)[] ellipse (1.8cm and 0.5cm);
        \draw [thick,blue](2.9,0)[] ellipse (1cm and 0.4cm);
        \draw [thick,violet](-0.9,0)[] ellipse (1cm and 0.4cm);
	\node[] () at (1.0,0.65)  {$\rm{a}$};		
	\node[] () at (3.0,0.65)  {$\rm{b}$};		
	\node[] () at (-1.0,0.65)  {$\rm{c}$};		
\end{tikzpicture}
\caption{A hypertree $\mcH$ with $V(\mcH)=\Set{1,2,3,4,5}, E(\mcH)=\Set{\rm{a},\rm{b},\rm{c}}$ and $\xi_{\mcH}(\rm{a})=\Set{1,2,3},\xi_{\mcH}(\rm{b})=\Set{3,4},\xi_{\mcH}(\rm{c})=\Set{1,5}$.}
\label{fig:ht}
\end{figure}
Note that a hypertree is a minimally connected hypergraph, but the reverse does not hold.\footnote{In graph theory, minimally connected graph and tree are equivalent.} Below is such an example.
\begin{Example} 
\label{ex:hypergraph}
The hypergraph in~\figref{fig:mch} is a minimally connected hypergraph but not a hypertree since $(1,\rm{a},2,\rm{b},3,\rm{c},1)$ is a cycle. 
\end{Example}

We can construct new hypergraphs from any given hypergraph $\mcH$ via the following operations. 
\begin{Definition}
\label{def:H/C}
For any subset $C\subseteq V(\mcH)$ of a hypergraph $\mcH$, $\mcH/ C$ is a hypergraph with 
\begin{align*}
V(\mcH/ C)&=V(\mcH)\setminus C, \\
\xi_{\mcH/ C}(\rm{e})&=\xi_{\mcH}(\rm{e})\setminus C \kern.5em\text{for} \kern.5em \rm{e}\in E(\mcH/ C),
\end{align*} 
where 
\begin{align*}
E(\mcH/ C)=\Set{ \rm{e}\in E(\mcH)\mid \xi_{\mcH}(\rm{e})\setminus C\neq `0}, 
\end{align*} 
%deleting is completely removing all edges invovling v, contracting is removing v.
%\begin{Definition}
%$\mcH_{C}$ is a hypergraph where the set of vertices is $V(\mcH_{C})=C$, the set of edges is 
%\begin{align*}
%E(\mcH_{C})=\Set{e\cap C\mid e\in E(\mcH), \xi_{\mcH}(e)\cap C\neq `0},
%\end{align*} 
%and the edge function is 
%\begin{align*}
%\xi_{\mcH_{C}}(e)=\xi_{\mcH}(e)\cap C, \kern.5em\text{for} \kern.5em e\in E(\mcH_C).
%\end{align*} 
%\end{Definition}
i.e., $\mcH/ C$ is a subhypergraph obtained from $\mcH$ by removing the vertices in $C$ from $V(\mcH)$, and then discarding the empty sets.
For notational simplicity, we use $\mcH_C$ to denote $\mcH/ (V(\mcH)\setminus C)$ and call it the subhypergraph of $\mcH$ induced by $C$.
\end{Definition}

Let $\Pi(V)$ be the set of partitions of $V(\mcH)$ into non-empty disjoint subsets.
\begin{Definition}
\label{def:H[P]}
For $\mcP\in \Pi(V)$, $\mcH[\mcP]$ is a hypergraph with 
\begin{align*}
V(\mcH[\mcP])&=\mcP, \\
\xi_{\mcH[\mcP]}(\rm{e})&=\Set{C\in \mcP\mid \xi_{\mcH}(\rm{e})\cap C\neq`0} \kern.3em\text{for} \kern.3em  \rm{e}\in E(\mcH[\mcP]),
\end{align*}
where $$E(\mcH[\mcP])=E(\mcH),$$
%\begin{align*}
%E(\mcH[\mcP])=\Bigg\{\bigg\{\bigcup_{C\in \mcP:C\cap\xi_{\mcH}(e)\neq`0}C\bigg\}\mid e\in E(\mcH)\Bigg\}, 
%\end{align*}
i.e., $\mcH[\mcP]$ is a hypergraph obtained from $\mcH$ by merging the vertices with respect to $\mcP$. 
\end{Definition}

\begin{figure}
	\centering
	\tikzstyle{dot}=[shape=circle,dashed,draw=gray!100,thick,inner sep=1pt,minimum size=11pt]
	\begin{subfigure}[b]{.45\textwidth}%setting to 0.9 for one column
		\centering
		\begin{tikzpicture}
		\node[dot] (4) at (0,0) {$4$};
		\node[dot] [right= 1.4cm of 4] (5) {$5$};
		\node[dot] [right= 1.4cm of 5] (6) {$6$};
    		\draw[thick,red] (0,0)[] ellipse (0.6cm and 0.4cm);
    		\draw[thick,blue] (1.8,0)[] ellipse (0.6cm and 0.4cm);
   		\draw[thick,violet] (3.6,0)[] ellipse (0.6cm and 0.4cm);
		\node[] () at (0,0.6)  {$\rm{a}$};
		\node[] () at (1.8,0.6)  {$\rm{b}$};
		\node[] () at (3.6,0.6)  {$\rm{c}$};
		\end{tikzpicture}
	\caption{$\mcH/C$ with $C=\Set{1,2,3}$.\label{fig:H/C}}
	\end{subfigure}
	\begin{subfigure}[b]{.45\textwidth}
		\centering
		\begin{tikzpicture}
		\node[dot] (2) at (0.95,0.2) {$2$};
		\node[dot] (3) at (2.5,-1.2) {$3$};
 		\node[dot] (1) at (-0.6,-1.2) {$1$};
   		\draw [thick,red,rotate around={40:(0,-0.4)}](0,-0.44) ellipse (1.44cm and 0.62cm);
    		\draw [thick,blue,rotate around={140:(1.9,-0.4)}](1.9,-0.38) ellipse (1.44cm and 0.62cm);
    		\draw[thick,violet] (1,-1.3) ellipse (2cm and 0.6cm);
		\node[] () at (-.7,0)  {$\rm{a}$};
		\node[] () at (2.65,0) {$\rm{b}$};
		\node[] () at (0.9,-2.05)  {$\rm{c}$};
		\end{tikzpicture}
	\caption{$\mcH_{C}$ with $C=\Set{1,2,3}$.\label{fig:Hc}}
	\end{subfigure}	
	\begin{subfigure}[b]{.45\textwidth}
		\centering
		\begin{tikzpicture}[scale=0.9]
		\node (1) at (0,0) {$1$};
		\node [right= 0.4cm of 1] (2) {$2$};
		\node [right= 0.4cm of 2] (3) {$3$};
		\node [right= 0.6cm of 3] (4) {$4$};
		\node [right= 0.4cm of 4] (5) {$5$};	
		\node[dot] [left= 0.6cm of 1] (6) {$6$};
    		\draw [thick,red] (2.0,0)[] ellipse (2.4cm and 0.6cm);
    		\draw [dashed,draw=gray!100,thick,inner sep=1pt,minimum size=11pt] (0.95,0) ellipse (1.1cm and 0.33cm);
    		\draw [thick,blue] (1.9,0)[] ellipse (2.6cm and 0.8cm);
    		\draw [dashed,draw=gray!100,thick,inner sep=1pt,minimum size=11pt] (3.45,0) ellipse (0.6cm and 0.3cm);
   		\draw [thick,violet] (0.5,0)[] ellipse (2cm and 0.6cm);
		\node[] () at (2.8,0.4)  {$\rm{a}$};		
		\node[] () at (3.3,0.9)  {$\rm{b}$};		
		\node[] () at (-1.3,0.44)  {$\rm{c}$};		
		\end{tikzpicture}	
		\caption{$\mcH[\mcP]$ with $\mcP=\Set{\Set{1,2,3},\Set{4,5},\Set{6}}$.\label{fig:H[P]}}
	\end{subfigure}
	\caption{Illustration of different hypergraph operations on $\mcH$ defined in~\figref{fig:mch}. (See Example~\ref{ex:hyp:oper}.)}
	\label{fig:h:op}
\end{figure}
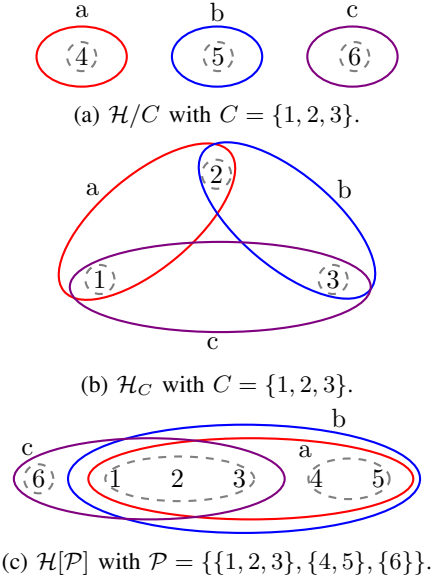

\begin{Example}
\label{ex:hyp:oper}
Consider the hypergraph $\mcH$ in~\figref{fig:mch} and let $C=\Set{1,2,3}$ and $\mcP=\Set{\Set{1,2,3},\Set{4,5},\Set{6}}$. \figref{fig:h:op} illustrates the different operations on the hypergraph defined above: $\mcH/ C$ is a hypergraph with 
\begin{align*}
&V(\mcH/ C)=\Set{4,5,6}, E(\mcH/ C)=\Set{\rm{a},\rm{b},\rm{c}},\\
&\xi_{\mcH/ C}(\rm{a})=\Set{4}, \xi_{\mcH/ C}(\rm{b})=\Set{5}, \xi_{\mcH/ C}(\rm{c})=\Set{6}.
\end{align*}
$\mcH_{C}$ is a hypergraph with 
\begin{align*}
&V(\mcH_{C})=\Set{1,2,3},E(\mcH_{C})=\Set{\rm{a},\rm{b},\rm{c}},\\ 
&\xi_{\mcH_{C}}(\rm{a})=\Set{1,2},\xi_{\mcH_{C}}(\rm{b})=\Set{2,3},\xi_{\mcH_{C}}(\rm{c})=\Set{1,3}.
\end{align*}  
$\mcH[\mcP]$ is a hypergraph with
\begin{align*}
&V(\mcH[\mcP])=\Set{\Set{1,2,3},\Set{4,5},\Set{6}}, E(\mcH[\mcP])=\Set{\rm{a},\rm{b},\rm{c}},\\ 
&\xi_{\mcH[\mcP]}(\rm{a})=\Set{1,2,3,4,5}, \xi_{\mcH[\mcP]}(\rm{b})=\Set{1,2,3,4,5},\\
&\xi_{\mcH[\mcP]}(\rm{c})=\Set{1,2,3,6}. \TheoremSymbol
\end{align*} 
\end{Example}

Finally, we shall introduce the notion of \emph{partition connectivity} for hypergraphs~\cite{chan10md}. Let $\Pi'(V)$ denote the set of all partitions of $V(\mcH)$ into at least two non-empty disjoint subsets,
i.e.,
\begin{align}
 \Pi'(V)=\Set{\mcP\in \Pi(V)\mid\abs{\mcP}>1}=\Pi(V)\setminus \Set{\Set{V}}.
\end{align}
\begin{Definition}
\label{def:par:conn}
With $\abs{V(\mcH)}\geq 2$, (which will be assumed hereafter), the \emph{partition connectivity} of a hypergraph $\mcH$ is defined as
\begin{subequations}
\label{eq:I(H)}
\begin{align}
I(\mcH)&:=\min_{\mcP\in \Pi'(V)}\frac1 {\abs{\mcP}-1}E_{\mcP}(\mcH), \quad \text{where} \label{eq:H:I}\\
\kern-1.5em E_{\mcP}(\mcH)\mkern-5mu&:=\mkern-5mu\sum_{C\in\mcP}\underbrace{|\Set{\rm{e}\mkern-3mu\in\mkern-3mu E(\mcH)\mkern-2mu\mid \xi_{\mcH}(\rm{e})\cap C\mkern-3mu\neq `0}|}_{=\mathtt{d}_{\mcH}(C)}\mkern-3mu-\mkern-15mu\underbrace{\abs{E(\mcH)}}_{=\mathtt{d}_{\mcH}(V(\mcH))}\kern-.5em, \kern-.5em \label{eq:H:IP}
\end{align}
\end{subequations}
which corresponds to the number of edges that cross the partition $\mcP$. 
\end{Definition}
Clearly, by the above definition, we have $I(\mcH)\geq 0$, with equality if and only if $\mcH$ is disconnected. 

The partition connectivity defined above stems from the multivariate mutual information (MMI) in~\cite{chan15mi}. More precisely, the MMI of $\RZ_V$ is defined as 
\begin{subequations}
	\label{eq:mmi}
\begin{align}
	I(\RZ_V) &:= \min_{\mcP\in\Pi'(V)}I_{\mcP}(\RZ_V), \quad\text{ where }\label{eq:I(Z_V)}\\
	I_{\mcP}(\RZ_V) &:=\frac{1}{|\mcP|-1}\left[\sum_{C\in\mcP}H(\RZ_C)-H(\RZ_V)\right].\label{eq:IP}
\end{align}
\end{subequations}
Then, assume that $\RZ_V$ is a hypergraphical source with respect to a hypergraph $\mcH$ and each edge corresponds to an independent bit, i.e., $H(\RX_{\rm{e}})=1$, for all $\rm{e}\in E(\mcH)$. It follows that 
\begin{align*}
H(\RZ_V)&=\mathtt{d}_{\mcH}(V(\mcH)),\\
H(\RZ_C)&=\mathtt{d}_{\mcH}(C),\quad \forall C\in \mcP\in\Pi'(V).
\end{align*}
With this, the MMI reduces to the partition connectivity defined in~\eqref{eq:I(H)}.
The MMI appeared as an upper bound on the unconstrained secrecy capacity in~\cite[eq. (26)]{csiszar04}. 
While it was shown in~\cite{chan2008tightness} the bound is not tight in general, it was also identified in~\cite{chan2008tightness} and~\cite{chan10md} to be tight in the important no-helper case. The problem studied in this paper is also this case. 
Therefore, we have 
\begin{align}
\label{eq:cs=mmi}
\CS(\infty)=I(\RZ_V).
\end{align}
The MMI was also called shared information in~\cite{narayan16}.
It was pointed out in~\cite[Lemma 5.1]{chan15mi} that the set of optimal solutions to~\eqref{eq:mmi} forms a lower semi-lattice with respect to the partial order ``$\preceq$'' on partitions defined as 
\begin{align}
	\mcP\preceq \mcP' \kern1em \text{iff}\kern1em \forall C\in \mcP, \ \exists C'\in \mcP':C\subseteq C', \label{eq:<P}
\end{align}
i.e., $\mcP$ \emph{is finer than} $\mcP'$ in the sense that $\mcP$ can be obtained from $\mcP'$ by further partitioning some parts of $\mcP'$. Hence, the set of optimal partitions to~\eqref{eq:I(H)}, denoted by~$\Uppi^*(\mcH)$, inherits the lattice structure as follows.

\begin{Proposition}[\mbox{\cite[Theorem~5.2]{chan15mi}}]
\label{pro:lattice:funda}
$\Uppi^*(\mcH)$ forms a lower semi-lattice with respect to the partial order~\eqref{eq:<P}. In particular, there is a unique finest optimal partition in $\Uppi^*(\mcH)$, denoted by $\mcP^*(\mcH)$ and referred to as the \emph{fundamental partition}.
\end{Proposition} 
Note, both $I(\mcH)$ and $\mcP^*(\mcH)$ can be computed in strongly polynomial time. Particularly, when $\mcH$ is disconnected, we have $I(\mcH)=0$ and $\mcP^*(\mcH)$ being the set of equivalent classes of $V(\mcH)$ under $\sim_{\mcH}$.

The fundamental partition has various properties and operational meanings. In particular, we will rely on the following property to derive our main results.
\begin{Proposition}[\mbox{\cite[Theorem~5.3]{chan15mi}}]
	\label{pro:P*}
	The fundamental partition $\mcP^*(\mcH)$ of a hypergraph $\mcH$ satisfies
	\begin{align}
	&\mcP^*(\mcH)\setminus \Set{\Set{v}\mid v \in V(\mcH)}\notag\\
	&=\op{maximal}\{C\subseteq V(\mcH)\mid \abs{C}>1,
	I(\mcH_{C})>I(\mcH)\}
	\end{align}
	where $\op{maximal}\mcF$ denotes the collection of inclusion-wise maximal sets in a set family $\mcF$, i.e., $\op{maximal} \mcF := \left\{ B \in \mcF \mid \not\exists B'\supsetneq B, B'\in \mcF \right\}$.
\end{Proposition}
The above has an elegant interpretation in data clustering~\cite{chan16cluster}: $\mcP^*(\mcH)$ is a clustering of the vertices in $\mcH$ such that the intra-cluster connectivity $I(\mcH_{C})$ for any non-singleton cluster $C\in\mcP^*(\mcH)$ is strictly larger than the inter-cluster connectivity $I(\mcH)$. 

We end this section with an example that illustrates the partition connectivity and the fundamental partition of hypergraphs. 
\begin{Example}
\label{ex:I(H):P*}
Consider the hypergraph in~\figref{fig:mch}. For $\mcP=\Set{\Set{1,2,3},\Set{4,5},\Set{6}}$, the expression in the definition~\eqref{eq:H:I} of $I(\mcH)$  is
\begin{align*}
\begin{split}
&\frac{1}{|\mcP|-1}E_{\mcP}(\mcH)\\
&=\frac{1}{3-1}\Big[\mathtt{d}_{\mcH}(\Set{1,2,3})+\mathtt{d}_{\mcH}(\Set{4,5})+\mathtt{d}_{\mcH}(\Set{6})\\
&\kern10.5em-\mathtt{d}_{\mcH}(\Set{1,2,3,4,5,6})\Big]\\
&=\frac{1}{3-1}[3+2+1-3]\\
&=1.5.
\end{split}
\end{align*}
The calculation for other partitions $\mcP$ can be done similarly. It can be checked that 
\begin{align*}
I(\mcH)&=1\text{ and }
\mcP^*(\mcH)=\Set{\Set{1,2,3},\Set{4},\Set{5},\Set{6}}.
\end{align*}
The non-singleton subsets with partition connectivity strictly larger than one are $\Set{1,2,3}$, which has the value
\begin{align*}
I\!\left(\mcH_{\Set{1,2,3}}\right)=1.5 \text{ with } \mcP^*\!\left(\mcH_{\Set{1,2,3}}\right)=\Set{\Set{1},\Set{2},\Set{3}}.
\end{align*}
$\mcH_{\Set{1,2,3}}$ is shown in~\figref{fig:Hc}. 
It turns out $\Set{1,2,3}$ is also the only non-singleton set in the fundamental partition as expected from Proposition~\ref{pro:P*}. 
\end{Example}
\section{Main Results}
\label{sec:results}
Unless otherwise stated, all the results in this section apply to a source $\RZ_V$ defined in Definition~\ref{def:mchs} that is hypergraphical with respect to a MCH $\mcH=(V,E,\xi)$.

First, we obtain an explicit formula for the unconstrained secrecy capacity $\CS(\infty)$.
\begin{Proposition}
	\label{pro:CS}
	The unconstrained secrecy capacity defined in~\eqref{eq:cs8} is
	\begin{align}
		\label{eq:MCH:CS}
	\CS(\infty)=\min_{\rm{e}\in E} w(\rm{e}).\TheoremSymbol
	\end{align}
\end{Proposition}
%Remark:
Although $\CS(\infty)$ has been characterized as a linear program in \cite{csiszar04} for general multiterminal sources, the above explicit characterization for MCH sources is new. 
%\begin{Proof}
%See Section~\ref{sec:proof:pro:4.1}.
%\end{Proof}
Indeed, the entire achievable rate region $\rsfsR$ can also be characterized explicitly as follows:
\begin{Theorem}
	\label{thm:MCH:Region}
	The achievable rate region $\rsfsR$ in~\eqref{eq:def:region} is
	\begin{align}
		\label{eq:MCH:R}
	\kern-.7em\rsfsR=\mkern-5mu\left \{\kern-.7em{\begin{array}{cc}{(\rK, r_V)\mkern-5mu\in\mathbb{R}_+^{|V|+1}}\mkern-2mu\left |{ \begin{array}{ll} \rK \le \CS(\infty) ,\\ r(B)\geq  [\kappa(\mcH/B)-1]\rK,\\ \forall B\subseteq C\in\mcP^*(\mcH)\end{array} }\right.\end{array}}\kern-1.1em\right \}\mkern-3mu, \mkern-10mu
	\end{align}
	where $\CS(\infty)$ is given by~\eqref{eq:MCH:CS},
	$\kappa(\mcH/ B)$ is the number of connected components of $\mcH/B$ defined in Definition~\ref{def:H/C} by removing the vertices in $B$, and $\mcP^*(\mcH)$ is the fundamental partition in Proposition~\ref{pro:lattice:funda}. It follows that private randomization at the users does not serve to reduce the public discussion rates nor increase the secret key rate. Furthermore, $(\rK, r_V)$ can be attained non-asymptotically with zero error and perfect secrecy through linear non-interactive discussion. In particular, it suffices to consider block length $n=1$ and the secret key can be chosen to be a function of an arbitrary edge random variable $\RX_{\rm{e}}, \rm{e}\in E$.
\end{Theorem}
%\begin{Proof}
%The converse and achievability proofs are in Section~\ref{sec:proof:them:4.1:con} and~\ref{sec:proof:them:4.1:ach}. 
%\end{Proof}
%Remarks: 
%We pause here to make some remarks: It is noteworthy that, in general, none of the inequalities in~\eqref{eq:MCH:R} is implied by the others, i.e., there is no redundancy in the specification of $\rsfsR$.\footnote{Some inequalities may be trivial when specialized to a particular MCH source.}  
It turns out that, for the MCH sources, 
zero error and perfect secrecy come at no additional cost in the public discussion rates. Furthermore, private randomization does not help increase the secret key rate nor decrease the discussion rates and can therefore be excluded at the outset to simplify the protocol. The achievability under block length $n=1$ suggests that the secret key can be generated sample by sample with no delay. We give a simple example below to illustrate
Propostition~\ref{pro:CS} and Theorem~\ref{thm:MCH:Region}.
\begin{Example}
\label{ex:mch:R}
Consider the MCH source in~Example~\ref{ex:mch}. As shown in Example~\ref{ex:I(H):P*}, $\mcP^*(\mcH)=\Set{\Set{1,2,3},\Set{4},\Set{5},\Set{6}}$. By~\eqref{eq:MCH:CS}, 
\begin{align*}
\CS(\infty)=\min\Set{H(\RX_{\rm{a}}),H(\RX_{\rm{b}}),H(\RX_{\rm{c}})}=H(\RX_{\rm{a}})=1,
\end{align*} 
and by~\eqref{eq:MCH:R},
	\begin{align*}
	\rsfsR=\left \{{\begin{array}{cc}{(\rK, r_V)\in\mathbb{R}_+^{7}}\left |{ \begin{array}{ll} \rK \le 1 ,\\ r_1+r_2\geq  \rK,\\ r_1+r_3\geq  \rK,\\ r_2+r_3\geq  \rK,\\ r_1+r_2+r_3\geq  2\rK \end{array} }\right.\end{array}}\right \}. 
	\end{align*}
In particular, the last inequality $r_1+r_2+r_3\geq 2\rK$ is because $\mcH/ \{1,2,3\}$, which is shown in~\figref{fig:H/C}, has $\kappa(\mcH/\{1,2,3\})=3$ connected components. The others constraints in~\eqref{eq:MCH:R} with $|B|=1$ are trivial and therefore omitted in the above expression.   
\end{Example}

Although the achievable rate region $\rsfsR$ is characterized explicitly by \eqref{eq:MCH:R}, its computation may still take exponential time as we go through all possible subsets $B$ and $C$ in the expression. Nevertheless, particularizing the above result to a hypergraphical source $\RZ_V$ where the corresponding hypergraph is a hypertree gives a simple characterization of $\rsfsR$ as follows.
\begin{Corollary}
\label{cor:HT}
For a source $\RZ_V$ hypergraphical with respect to a hypertree $\mcH$ (see Definition~\ref{def:HT}), we have
	\begin{align}
	\label{eq:HT:region}
	\rsfsR=\left \{{\begin{array}{cc}{(\rK, r_V)\in\mathbb{R}_+^{|V|+1}}\left |{ \begin{array}{ll} \rK \le \CS(\infty) ,\\ r_i\geq  [\mathtt{d}_{\mcH}(i)-1]\rK,\\ \forall i\in V\end{array} }\right.\end{array}}\kern-.7em \right \}, 
	\end{align}
	where $\CS(\infty)$ is as defined in~\eqref{eq:MCH:CS} and $\mathtt{d}_{\mcH}(i)$ is the degree of vertex $i$ in the hypertree $\mcH$ defined in~\eqref{eq:deg:v}.
\end{Corollary}
%\begin{Proof}
%See Appendix~\ref{sec:proof:source}.
%\end{Proof}
Our result generalizes~\cite[Theorem 4.2]{chan17isit}, which is the special case when the hypertree is a tree. 
\begin{Example}
With $V=\Set{1,2,3,4,5}$, define $\RZ_V$ as
\begin{equation*}
	\begin{aligned}
		\RZ_1&:=(\RX_{\rm{a}},\RX_{\rm{c}}),&   
		\RZ_2&:=\RX_{\rm{a}},&
		\RZ_3&:=(\RX_{\rm{a}},\RX_{\rm{b}}),&\\
		\RZ_4&:=\RX_{\rm{b}},&
		\RZ_5&:=\RX_{\rm{c}},&
	\end{aligned}
\end{equation*}
where $\RX_{i}$'s are independent with $H(\RX_{\rm{a}})=2$ and $H(\RX_{\rm{b}})=H(\RX_{\rm{c}})=1$. This is a hypergraphical source with respect to the hypertree in~\figref{fig:ht} with weight $w(\rm{a})=H(\RX_{\rm{a}})=2$, $w(\rm{b})=H(\RX_{\rm{b}})=1$, and $w(\rm{c})=H(\RX_{\rm{c}})=1$. By~\eqref{eq:MCH:CS} and~\eqref{eq:HT:region}, we have
	\begin{align*}
	\rsfsR=\left \{{\begin{array}{cc}{(\rK, r_V)\in\mathbb{R}_+^{6}}\left |{ \begin{array}{ll} \rK \le 1 ,\\ r_1\geq  \rK,\\ r_3\geq  \rK \end{array} }\right.\end{array}}\right \} 
	\end{align*} 
because $\kappa(\mcH/\{i\})$ equals $2$ for $i\in\{1,3\}$ and $1$ for $i\in\{2,4,5\}$.
This result is not covered by~\cite[Theorem 4.2]{chan17isit} because $\mcH$ is not a tree.
\end{Example}

Despite the above result, the computation of $\rsfsR$ for the general MCH sources may require a lot of machinery as mentioned above. Fortunately, for the constrained secrecy capacity $\CS(R)$, we obtain a closed-form formula that is easy to compute. 
\begin{Theorem}
\label{thm:MCH:CS(R)}	
	The constrained secrecy capacity $\CS(R)$ defined in~\eqref{CS(R)} is
	\begin{align}	
		\label{eq:MCH:CS(R)}
	\CS(R)=\min\left\{\frac{R}{\abs{E}-1},\CS(\infty) \right\},
	\end{align}
	where $\CS(\infty)$ is given by~\eqref{eq:MCH:CS}.
\end{Theorem}
%\begin{Proof}
%See Appendix~\ref{sec:proof:source}.
%\end{Proof}
Observe that the optimal trade-off is characterized simply by the number of edges. By equating the two terms in the minimization in~\eqref{eq:MCH:CS(R)}, we obtain the following formula for the communication complexity.
\begin{Corollary}
\label{cor:MCH:RS}
The communication complexity defined in~\eqref{eq:CS} is
\begin{align}
	\label{eq:MCH:RS}
\RS=[\abs{E}-1]\CS(\infty),
\end{align}
where $\CS(\infty)$ is as defined in~\eqref{eq:MCH:CS}.
\end{Corollary}

The following example illustrates the results of Theorem~\ref{thm:MCH:CS(R)} and Corollary~\ref{cor:MCH:RS}.  
\begin{Example} 
Consider the MCH source defined in Example~\ref{ex:mch}. According to~\eqref{eq:MCH:CS},~\eqref{eq:MCH:CS(R)} and~\eqref{eq:MCH:RS}, we have 
%\begin{align*}
$\CS(R)=\min\left\{\frac{R}{2},1\right\}$ and $\RS=2$ since there are $|E|=3$ edges and the minimum weight is $\min_{\rm{e}\in E}w(\rm{e})=1$.
%\end{align*}
\end{Example}

\section{Proofs}
\label{sec:results:proof}
%Toward proving the main results, we clarify some combinatorial properties of the hypertree and the minimally connected hypergraph (MCH). These properties may be of independent interest from a purely hypergraph theoretic viewpoint. We first establish the following alternative characterization of the hypertree through the partition connectivity and fundamental partition.  
\subsection{Proof of Proposition~\ref{pro:CS}}
\label{sec:proof:pro:4.1}
%\begin{Proof}
Recall that $\CS(\infty)=I(\RZ_V)$ in~\eqref{eq:cs=mmi}. To show the achievability ``$\geq$'' of~\eqref{eq:MCH:CS}, it suffices to show that 
\begin{align*}
I_{\mcP}(\RZ_V)\geq \min_{\rm{e}\in E(\mcH)} w(\rm{e}), \quad \forall \mcP\in \Pi'(V).
\end{align*}
To that end, consider any $\mcP\in \Pi'(V)$ and let $q=|\mcP|$. Since $\mcH$ is connected, we can always enumerate $\mcP$ as $\left\{C_1,\dots, C_q\right\}$ such that $C_i$ and $\bigcup_{j=i+1}^q C_j$ share at least one edge for all $1\leq i\leq q-1$, i.e.,
\begin{multline}
\label{eq:C:e>=1}
\forall\, 1\leq i\leq q-1, \exists\, \rm{e}\in E(\mcH) \\ \text{ s.t. }\xi_{\mcH}(\rm{e})\cap C_i\neq \emptyset, \xi_{\mcH}(\rm{e})\cap\bigcup_{j=i+1}^q C_j \neq \emptyset.
\end{multline}  
This can be done via reordering in the following way: Let $\mcP = \{C_1, \ldots, C_q\}$ be an arbitrary enumeration of the elements in the partition. Now, we are going to construct a permutation $\pi: [q]\to [q]$ to reorder $\mcP$ such that it satisfies~\eqref{eq:C:e>=1}. First, define $$\pi(q):=q.$$
Then, for $i$ from $2$ to $q$, pick an element $C_\ell\in\mcP\setminus \left\{C_{\pi^{-1}(q-i+2)},\dots,C_{\pi^{-1}(q)}\right\}$ such that it shares at least one edge with the set of all the previous picked elements $\bigcup_{j=q-i+2}^q C_{\pi^{-1}(j)}$. For each $i$, there always exists at least one such element, otherwise, $\mcH$ becomes disconnected. Define $$\pi(\ell):=q-i+1.$$
By construction, the reordered $C_{\pi^{-1}(1)},\dots,C_{\pi^{-1}(q)}$ satisfies the desired property~\eqref{eq:C:e>=1}. 

Now, assuming \eqref{eq:C:e>=1} holds, and upon expanding $I_{\mcP}(\RZ_V)$ in terms of Shannon's mutual information~\cite[eq. (5.18)]{chan15mi}, we have
\begin{align*}
	I_{\mcP}(\RZ_V)&=\frac1{q-1}\sum_{i=1}^{q-1} I\left(\RZ_{C_i} \wedge \RZ_{\bigcup_{j=i+1}^q C_j}\right)\\
				 &\geq \frac1{q-1}\sum_{i=1}^{q-1}\min_{\rm{e}\in E(\mcH)} w(\rm{e})\\
				 &=\min_{\rm{e}\in E(\mcH)} w(\rm{e})
\end{align*}
as desired. Here, the inequality follows from~\eqref{eq:C:e>=1}. 

To prove the converse, let $\rm{e}^*$ be the optimal solution to the R.H.S. of~\eqref{eq:MCH:CS}. Let $\mcP'$ be the set of equivalent classes of $\mcH$ after removing edge $\rm{e}^*$. It follows that $\mcP'\in \Pi'(V)$ due to the assumption that $\mcH$ is minimally connected. Then, 
\begin{align*}
\CS(\infty)&=I(\RZ_V)\\
&\utag{a}\leq I_{\mcP'}(\RZ_V) \\
&\utag{b}=\frac{1}{|\mcP'|-1}\left[\sum_{C\in \mcP'}H(\RX_{\rm{e}^*})-H(\RX_{\rm{e}^*})\right]\\
&\utag{c}=w(\rm{e}^*)
\end{align*} 
where~\uref{a} follows from~\eqref{eq:I(Z_V)}; \uref{b} is because of the independence of the edge random variables and the fact that $\rm{e}^*$ is the only edge that crosses $\mcP'$; \uref{c} follows from~\eqref{eq:hyper:weight}. Therefore, we have proved the converse ``$\leq$'' of~\eqref{eq:MCH:CS}, and thereby Proposition~\ref{pro:CS}. 
%\end{Proof}

\subsection{Proof of Theorem~\ref{thm:MCH:Region}: Converse}
\label{sec:proof:them:4.1:con}
%\begin{Proof}
The proof will make use of the following technical result in~\cite{chan17isit}, which provides an outer bound on the achievable rate region $\rsfsR$ for a general multiterminal source. 
\begin{Proposition}[\mbox{\cite[Theorem~4.1]{chan17isit}}]
	\label{pro:LB}
	For any $(\rK,r_V)\in\rsfsR$, we have 
	\begin{align}
		\label{eq:LB}
		 r(B)&\geq (\abs{\mcP}-1)[\rK-I_{\mcP}(\RZ_{V\setminus B})]
	\end{align}
	for any $B\subseteq V$ with size $\abs{B}<|V|-1$ and $\mcP\in\Pi'(V\setminus B)$, where $I_{\mcP}$ is defined in~\eqref{eq:IP}. 
\end{Proposition}
Our converse part is obtained by specializing the above outer bound to the MCH sources. 
However, instead of applying~\eqref{eq:LB} for all $B$ with $|B|<|V|-1$ and $\mcP\in\Pi'(V\setminus B)$, it suffices to consider only those $B\subseteq C\in\mcP^*(\mcH)$ with $\kappa(\mcH/B)>1$ and $\mcP$ being the fundamental partition $\mcP^*(\mcH/B)$ of hypergraph $\mcH/B$. Indeed, we show in Appendix~\ref{sec:proof:converse:alternative} that
\begin{itemize}
\item~\eqref{eq:LB} is trivial for all $B$ with $\kappa(\mcH/B)=1$;
\item~\eqref{eq:LB} is redundant for all $\mcP\in \Pi'(V\setminus B)\setminus\Set{\mcP^*(\mcH/B)}$;
\item~\eqref{eq:LB} is redundant for all $B$ with $B\not\subseteq C\,\,\, \forall C\in\mcP^*(\mcH)$. 
\end{itemize}
Therefore, the above restriction does not lose any optimality and provides a concise characterization of the achievable rate region. 

Consider any $B\subseteq C\in\mcP^*(\mcH)$. Since $\kappa(\mcH/B)$ is a positive integer, we have the following two cases. 

\underline{Case 1:} $\kappa(\mcH/B)= 1$, i.e., the hypergraph $\mcH/B$ is connected. Then,
\begin{align*}
r(B)\geq [\kappa(\mcH/B)-1]\rK=0
\end{align*} 
holds trivially. 

\underline{Case 2:} $\kappa(\mcH/B)>1$, i.e., the hypergraph $\mcH/B$ is disconnected. 
Let $\mcP=\mcP^*(\mcH/B)$, namely the set of equivalent classes of hypergraph $\mcH/B$. 
It follows that  
\begin{align*}
\mcP \in\Pi'(V\setminus B)  \kern1em\text{and}\kern1em \abs{\mcP}=\kappa(\mcH/B). 
\end{align*}
For such $\mcP$, we have 
\begin{align*}
I_{\mcP}(\RZ_{V\setminus B})=0
\end{align*}
because, by the definition of $\mcP$, every hyperedge of the corresponding hypergraph $\mcH/B$ of $\RZ_{V\setminus B}$ is entirely contained by a part of $\mcP$, i.e., 
\begin{align*}
\forall\rm{e}\in E(\mcH/B), \exists C\in \mcP: \xi_{\mcH/B}(\rm{e})\subseteq C.
\end{align*} 
In other words, no edges cross $\mcP$. Now, applying the lower bound~\eqref{eq:LB} with the partition $\mcP$, we get 
\begin{align*}
r(B)&\geq (\abs{\mcP}-1)[\rK-I_{\mcP}(V\setminus B)]\\
	&=(\kappa(\mcH/B)-1)\rK 
\end{align*}
This, together with the fact that $\rK\leq \CS(\infty)$, completes the converse proof of Theorem~\ref{thm:MCH:Region}.
%\end{proof}

\subsection{Proof of Theorem~\ref{thm:MCH:Region}: Achievability}
\label{sec:proof:them:4.1:ach}

Before presenting the proof, let us give some technical results that constitute the basic ingredients of the proof.
 \begin{Lemma}
	\label{lem:MCH}
	For any MCH $\mcH$, we have
\begin{enumerate}[label=(\roman*)]
\item $\mcH$ is a MCH iff $\mcH[\mcP^*(\mcH)]$ (see Definition~\ref{def:H[P]}) is a hypertree. \label{mch-to-hypertree}
\item Furthermore, 
\begin{align}
	\label{eq:mch:C(H)=d(C)}
&\kappa(\mcH/C)=\mathtt{d}_{\mcH}(C),\quad\forall C\in \mcP^*(\mcH),%\kern1em \text{and}
\end{align} 
and, in particular, 
\begin{align}
	\label{eq:mch:d(C)=E}
&\sum_{C\in\mcP^*(\mcH)}\left[\mathtt{d}_{\mcH}(C)-1\right]=\abs{E(\mcH)}-1,
\end{align}
where $\mathtt{d}_{\mcH}(C)$ is the degree of $C$ in $\mcH$ defined in~\eqref{eq:deg:c}.\label{degree:cc:edges}
\end{enumerate}
\end{Lemma}
%Remark: 
We remark that equation~\eqref{eq:mch:d(C)=E} holds not only for MCH. It continues to hold even if $\mcH$ has self-contained edges contained completely within some $C\in\mcP^{*}(\mcH)$, i.e., $\exists \rm{e}\in E(\mcH)$ s.t. $\xi_{\mcH}(e)\subseteq C$ for some $C\in\mcP^*(\mcH)$, as long as its induced hypergraph $\mcH[\mcP^*(\mcH)]$ is connected and cycle-free. In other words, $\mcH[\mcP^*(\mcH)]$ may contain loops. For example, for the hypergraph in~\figref{fig:h:loop}, which is not minimally connected,~\eqref{eq:mch:d(C)=E} also holds. 
\begin{Proof}
See Appendix~\ref{sec:proof:lem:MCH}.
%See the full version~\cite{chan18isit} of the paper.
\end{Proof} 
The first assertion in the above lemma provides an alternative characterization of the MCH. More importantly, it elucidates that the paths between distinct $C\in\mcP^*(\mcH)$ are unique, which naturally suggests an optimal achieving scheme that propagates the secret key along those paths, similar to the tree-packing protocol for the PIN model in~\cite{nitinawarat-ye10,nitinawarat10}. 
The second assertion establishes a relationship between the degree, the number of connected components, and the number of edges in MCH. With this, we can characterize the total amount discussion by all the users or by users in each $C\in\mcP^*(\mcH)$ in our proposed scheme. We give an example below to illustrate the properties of the MCH in Lemma~\ref{lem:MCH}.

\begin{Example}
Let us consider the MCH $\mcH$ in~\figref{fig:mch}. Recall that $\mcP^*(\mcH)=\Set{\Set{1,2,3},\Set{4},\Set{5},\Set{6}}$. Then, by Definition~\ref{def:H[P]}, $\mcH[\mcP^*(\mcH)]$ is the hypergraph shown in~\figref{fig:h[p*]} with vertex set 
\begin{align*}
V(\mcH[\mcP^*(\mcH)])=\Set{\Set{1,2,3},\Set{4},\Set{5},\Set{6}},
\end{align*} 
edge set $$E(\mcH[\mcP^*(\mcH)])=\Set{\rm{a},\rm{b},\rm{c}},$$ and edge function 
\begin{align*}
&\xi_{\mcH[\mcP^*(\mcH)]}(\rm{a})=\Set{1,2,3,4},\\ 
&\xi_{\mcH[\mcP^*(\mcH)]}(\rm{b})=\Set{1,2,3,5},\\
&\xi_{\mcH[\mcP^*(\mcH)]}(\rm{c})=\Set{1,2,3,6}.
\end{align*}
$\mcH[\mcP^*(\mcH)]$ is indeed a hypertree, as substantiated by Lemma~\ref{lem:MCH}~\ref{mch-to-hypertree}.
In contrast, see~\figref{fig:H[P]} for $\mcH[\mcP]$ with $\mcP=\{\{1,2,3\},\{4,5\},\{6\}\}$, which is not a hypertree.   
It can be readily verified that
\begin{align*}
\kappa(\mcH/C)&=\mathtt{d}_{\mcH}(C)=\begin{cases}
					3,& C=\Set{1,2,3},\\
					1,& C=\Set{i},  i\in\Set{4,5,6} 
				    \end{cases}
\end{align*}
as expected by~\eqref{eq:mch:C(H)=d(C)}.
It then immediately follows that $$\sum_{C\in\mcP^*(\mcH)}\left[\mathtt{d}_{\mcH}(C)-1\right]=2,$$ which equals $\abs{E(\mcH)}-1$ as desired by~\eqref{eq:mch:d(C)=E}.
\end{Example}
\begin{figure}
\centering
\tikzstyle{dot}=[shape=circle,dashed,draw=gray!100,thick,inner sep=1pt,minimum size=11pt]
\begin{tikzpicture}[scale=0.8]
	\node (1) at (0.06,0) {$1$};
	\node [right= 0.06cm of 1] (2) {$2$};
	\node [right= 0.06cm of 2] (3) {$3$};
	\node[dot] [above left=0.5cm and 0.5cm of 1] (4) {$4$};
	\node[dot] [above right=0.5cm and 0.5cm of 3] (5) {$5$};
	\node[dot] [below= 1cm of 2] (6) {$6$};
    \draw [thick,red,rotate around={-30:(-0.1,0.7)}](-0.1,0.7) ellipse (2cm and 0.9cm);
    \draw [thick,blue,rotate around={30:(1.6,0.7)}](1.6,0.7) ellipse (2cm and 0.9cm);
    \draw [thick,violet,rotate around={90:(0.8,-1)}](0.8,-1) ellipse (2cm and 1.1cm);
    \draw [dashed,draw=gray!100,thick,inner sep=1pt,minimum size=11pt] (0.7,0) ellipse (0.8cm and 0.33cm);
    	\node[] () at (-1.75,0.65)  {$\rm{a}$};		
    	\node[] () at (3.3,0.65)  {$\rm{b}$};		
    	\node[] () at (-0.5,-1.5)  {$\rm{c}$};		
\end{tikzpicture}
	\caption{$\mcH[\mcP^*(\mcH)]$ for $\mcH$ defined in~\figref{fig:mch}.}
	\label{fig:h[p*]}
\end{figure}

Now, we proceed to understand the individual discussion rate within each $C\in\mcP^*(\mcH)$. To this end, we will rely on the following technical results. To proceed, 
consider any $C\in\mcP^*(\mcH)$. Define
\begin{subequations}
\label{def:eq:H_Ec}
\begin{align}
E_C:=\left\{\rm{e}\in E(\mcH)\mid \xi_\mcH(\rm{e})\cap C\neq\emptyset\right\}
\end{align}
as the collection of edges incident on $C$, and 
\begin{align}
V_C:=\left\{i\in V(\mcH)\mid \rm{e}\in E_C, i\in \xi_\mcH(\rm{e})\right\}
\end{align}
as the collection of vertices incident on some edges in $E_C$. %\xi_{\mcH_{E_C}}(e)=\xi_{\mcH}(e),\forall e\in E_C
Then, let
\begin{align}
\mcH_{E_C}:=\left(V_C,E_C,\xi\right)
\end{align}
\end{subequations}
denotes the subhypergraph of $\mcH$ induced by $E_C$. The following simple observation pertaining to~$\mcH_{E_C}$~will be useful in analyzing the discussion within each $C$.
\begin{Lemma}
	\label{lem:H_Ec}
For MCH $\mcH$ and all $C\in\mcP^*(\mcH)$ of size $|C|>1$, $\mcH_{E_C}$ defined above in~\eqref{def:eq:H_Ec} is minimally connected and satisfies the following properties:
\begin{enumerate}[label=(\roman*)]
\item $\forall i \in V_C, \mathtt{d}_{\mcH_{E_C}}(i)\geq 1$, with equality iff $i\in V_C\setminus C$; \label{H_Ec:i}
\item $\forall e\in E_C,\, \exists i,j \in \xi_{\mcH_{E_C}}(\rm{e})$ s.t. $\mathtt{d}_{\mcH_{E_C}}(i)=1<\mathtt{d}_{\mcH_{E_C}}(j)$.\label{H_Ec:ii}
\end{enumerate}
\end{Lemma}
%Remark: For $|C|=1$, the above properties except the only if part of~\ref{H_Ec:i} also hold. 
\begin{Proof}
See Appendix~\ref{sec:proof:lem:H_Ec}.
\end{Proof}
The following example helps illustrate the above properties.
\begin{Example}
Let us consider the MCH $\mcH$ in~\figref{fig:H_Ec:H}. It can be verified that 
\begin{align*}
\mcP^*(\mcH)=\Set{\Set{1,2},\Set{3,4,8},\Set{5},\Set{6},\Set{7},\Set{9}}.
\end{align*}
For $C=\Set{1,2}\in\mcP^*(\mcH)$, by~\eqref{def:eq:H_Ec},%For such $C$, 
\begin{align*}
E_C=\Set{\rm{a},\rm{b},\rm{c}}, \quad V_C=\Set{1,2,3,5,6},
\end{align*}
and $\mcH_{E_C}$ is shown in~\figref{fig:H_Ec}, which is minimally connected. 
From~\figref{fig:H_Ec}, it is easy to see that
\begin{align*}
 \mathtt{d}_{\mcH_{E_C}}(i)=\begin{cases}
					1,& i\in\Set{3,5,6},\\ 
					2,& i=1,\\
					3,& i=2
				    \end{cases}
\end{align*}
For edge $\rm{a}$, 
\begin{align*}
\xi_{\mcH_{E_C}}(\rm{a})=\Set{1,2,5} \text{ and } \mathtt{d}_{\mcH_{E_C}}(5)=1<\mathtt{d}_{\mcH_{E_C}}(1).
\end{align*}
 For edge $\rm{b}$, 
 \begin{align*}
\xi_{\mcH_{E_C}}(\rm{b})=\Set{1,2,6} \text{ and } \mathtt{d}_{\mcH_{E_C}}(6)=1<\mathtt{d}_{\mcH_{E_C}}(2).
\end{align*}
 For edge $\rm{c}$,
\begin{align*}
\xi_{\mcH_{E_C}}(\rm{c})=\Set{2,3} \text{ and } \mathtt{d}_{\mcH_{E_C}}(3)=1<\mathtt{d}_{\mcH_{E_C}}(2).
\end{align*}
Therefore, the two properties are satisfied, as substantiated by the above lemma.
\end{Example}
\begin{figure}
	\centering
	\tikzstyle{dot}=[shape=circle,dashed,draw=gray!100,thick,inner sep=1pt,minimum size=11pt]
	\begin{subfigure}[b]{.45\textwidth}
		\centering
		{\begin{tikzpicture}
			\node[dot] (2) at (-0.1,0) {$2$};
			\node[dot] (3) at (1.6,0) {$3$};
			\node[dot] (4) at (2.6,0) {$4$};
			\node[dot] (1) at (-1.1,0) {$1$};
			\node[dot] (5) at (-0.6,1.5) {$5$};
			\node[dot] (6) at (-0.6,-1.5) {$6$};
			\node[dot] (7) at (2.1,1.5) {$7$};
			\node[dot] (8) at (2.1,-1.5) {$9$};
			\node[dot] (8) at (2.1,-.55) {$8$};
       			 \draw [thick,red,rotate around={90:(-0.6,0.6)}](-0.6,0.6) ellipse (1.4cm and .85cm);
       			 \draw [thick,blue,rotate around={90:(-0.6,-0.6)}](-0.6,-0.6) ellipse (1.4cm and .85cm);
        			\draw [thick,violet,rotate around={90:(2.1,0.6)}](2.1,0.6) ellipse (1.4cm and .85cm);
        			\draw [thick,cyan,rotate around={90:(2.1,-0.6)}](2.1,-0.6) ellipse (1.4cm and .85cm);
       			 \draw [thick,](0.75,0)[] ellipse (1.1cm and 0.5cm);
			 \node[] () at (-1.6,1)  {$\rm{a}$};		
			\node[] () at (-1.6,-1)  {$\rm{b}$};		
			\node[] () at (0.75,0.65)  {$\rm{c}$};		
			\node[] () at (3.1,1)  {$\rm{d}$};		
			\node[] () at (3.1,-1)  {$\rm{e}$};
		\end{tikzpicture}
		}
	\caption{Hypergraph $\mcH$ with $V(\mcH)=\Set{1,2,3,4,5,6,7,8,9}, E(\mcH)=\Set{\rm{a},\rm{b},\rm{c},\rm{d},\rm{e}}$ and $\xi_{\mcH}(\rm{a})=\Set{1,2,5},\xi_{\mcH}(\rm{b})=\Set{1,2,6}, \xi_{\mcH}(\rm{c})=\Set{2,3}$, $\xi_{\mcH}(\rm{d})=\Set{3,4,7,8},\xi_{\mcH}(\rm{e})=\Set{3,4,8,9}$.}
\label{fig:H_Ec:H}
	\end{subfigure}
	\begin{subfigure}[b]{.45\textwidth}
		\centering
		{\begin{tikzpicture}
			\node[dot] (2) at (-0.1,0) {$2$};
			\node[dot] (3) at (1.6,0) {$3$};
			\node[dot] (1) at (-1.1,0) {$1$};
			\node[dot] (5) at (-0.6,1.5) {$5$};
			\node[dot] (6) at (-0.6,-1.5) {$6$};
       			\draw [thick,red,rotate around={90:(-0.6,0.6)}](-0.6,0.6) ellipse (1.4cm and .85cm);
       			\draw [thick,blue,rotate around={90:(-0.6,-0.6)}](-0.6,-0.6) ellipse (1.4cm and .85cm);
        			\draw [thick,](0.75,0)[] ellipse (1.1cm and 0.5cm);
			\node[] () at (-1.6,1)  {$\rm{a}$};		
			\node[] () at (-1.6,-1)  {$\rm{b}$};		
			\node[] () at (0.75,0.65)  {$\rm{c}$};		
		\end{tikzpicture}
		}
		\caption{Hypergraph $\mcH_{E_C}$ with $C=\Set{1,2}$ for $\mcH$ defined in~\figref{fig:H_Ec:H}.%, E_C=\Set{\rm{a},\rm{b},\rm{c}}, V_{C}=\Set{1,2,3,5,6}$, and $\xi_{\mcH}(\rm{a})=\Set{1,2,5},\xi_{\mcH}(\rm{b})=\Set{1,2,6}, \xi_{\mcH}(\rm{c})=\Set{2,3}$.
		}
		\label{fig:H_Ec}
	\end{subfigure}
\caption{Illustration of subhypergraph $\mcH_{E_C}$ of $\mcH$.}
\end{figure}

A major step toward understanding the individual discussion rate within each $C\in\mcP^*(\mcH)$ is to exploit a combinatorial property of $\kappa(\mcH/B), B\subseteq C$ stated below.
\begin{Lemma}
	\label{lem:supermodular}
For any $C\in\mcP^*(\mcH)$ of a MCH $\mcH$, we have for all $S, T\subseteq C$,
\begin{align}
\label{eq:supermodular}
\kern-.5em\kappa(\mcH/S)+\kappa(\mcH/T)\mkern-2mu\leq \kappa(\mcH/(S\cup T))+\kappa(\mcH/(S\cap T)),
\end{align}
i.e., $\kappa(\mcH/B)$ is supermodular in $B\subseteq C$. 	
\end{Lemma}
\begin{Proof}
See Appendix~\ref{sec:proof:lem:supermodular}.
\end{Proof}
\begin{Example}
Consider the MCH $\mcH$ in~\figref{fig:H_Ec:H}. Let $C=\Set{3,4,8}\in\mcP^*(\mcH)$. Set $S=\Set{3,4}$ and $T=\Set{4,8}$. It is easy to see that $\kappa(\mcH/S)=2$, $\kappa(\mcH/T)=\kappa(\mcH/(S\cap T))=1$, and $\kappa(\mcH/(S\cup T))=3$. Therefore, we have~\eqref{eq:supermodular} holds with strict inequality.
\end{Example}
\begin{Example}
Consider the MCH $\mcH$ in~\figref{fig:mch}. Let $C=\Set{1,2,3}\in\mcP^*(\mcH)$. For $S=\Set{1,2}$, and $T=\Set{2,3}$. It is easy to see that $\kappa(\mcH/S)=\kappa(\mcH/T)=2, \kappa(\mcH/(S\cup T))=3$, and $\kappa(\mcH/(S\cap T))=1$. Therefore, we have~\eqref{eq:supermodular} holds with equality.
\end{Example}

For each $C\in\mcP^*(\mcH)$, define  
	\begin{align*}
		\label{eq:Rc}
	\mathscr{R}_C:=\left\{\left. r_C\in\mathbb{R}_+^{|C|} \right|  r(B)\geq [\kappa(\mcH/B)-1]\rK,\forall B\subseteq C \right\}.
	\end{align*}
Upon using the above two lemmas, we find that $\mathscr{R}_C$ forms a special kind of polyhedron, i.e., a \emph{contra-polymatroid} in the terminology of matroid theory (see, e.g.,~\cite{Edmonds,Oxley}). Certain rate regions of the multiple access channel~\cite{tse98} and distributed source coding problems~\cite{junchen04} are also known to have this specific combinatorial structure.

To see this, define
\begin{align}
f(B):=[\kappa(\mcH/B)-1]\rK,\quad\forall B\subseteq C\in\mcP^*(\mcH),
\end{align} 
where $\rK\geq 0$.
Then, the set function $f:2^C\to `R_+$ satisfies the following properties:
\begin{enumerate}
\item $f(\emptyset)=0$   (normalized) \label{property:nor}
\item $f(S)\leq f(T), \forall S\subseteq T$ (nondecreasing)
\item $f(S)+f(T)\leq f(S\cup T)+f(S\cap T)$ (supermodular).
\end{enumerate}
Property 1) holds by definition; 2) follows from \ref{H_Ec:ii} of Lemma~\ref{lem:H_Ec}; and 3) follows from Lemma~\ref{lem:supermodular}. 
By definition of contra-polymatroids~\cite{Edmonds,Oxley}, we conclude that $\mathscr{R}_C$ is a contra-polymatroid. One of the key properties of this combinatorial structure is that we can exactly characterize all the extreme points. This has been pointed out in~\cite{Edmonds}. For completeness, below is a specialization of this property customized to the setting in this work. 
%Without loss of generality, we assume that $C=\left\{1,\dots,\abs{C}\right\}$. 
\begin{Corollary}[\mbox{\cite{Edmonds}, \cite[Lemma~3.3]{tse98}}] 
	\label{lem:extreme_point}
For each $\mathscr{R}_C, C=\left\{i_1,i_2,\dots,i_{|C|}\right\}\in\mcP^*(\mcH)$, we have
\begin{enumerate}[label=(\roman*)]
\item an $\hat{r}_C=\left(\hat{r}_{i_1},\hat{r}_{i_2},\dots,\hat{r}_{i_{|C|}}\right)\in \mathscr{R}_C$ is an extreme point of $\mathscr{R}_C$ iff $\hat{r}_C$ can be expressed as
\begin{subequations}
\begin{align}
\hat{r}_{\pi(i_1)}&=[\kappa(\mcH/\{\pi(i_1)\})-1]\rK,\\
\begin{split}
\hat{r}_{\pi(i_{\ell})}&=[\kappa(\mcH/\Set{\pi(i_1),\dots,\pi(i_{\ell})}) \\ &\kern3em-\kappa(\mcH/\Set{\pi(i_1),\dots,\pi(i_{\ell-1})})]\rK, 
\end{split}
\end{align}
\end{subequations}
for $\ell=2,\dots, |C|$, where $\pi(i_1),\pi(i_2),\dots,\pi(i_{|C|})$ is a permutation of $i_1,i_2,\dots,i_{|C|}$. \label{pp:Pc:extreme_point}
\item Furthermore, any point of $\mathscr{R}_C$ is dominated by some convex combination of these extreme points. Here, $r_C$ is said to be dominated by $r'_C$, indicated by $r_C\geq r'_C$, if $r_{i_\ell}\geq r^{\prime}_{i_\ell}$ for all $\ell=1,\dots,|C|$.\label{pp:Pc:internal_point}
\end{enumerate}
\end{Corollary}
\begin{Proof}
See Appendix~\ref{sec:proof:lem:extreme_point}.
\end{Proof}
For each $\mathscr{R}_C, C\in\mcP^*(\mcH)$, we see from Corollary~\ref{lem:extreme_point} that the number of extreme points can be $\abs{C}!$ (These extreme points may not be distinct), because the number of distinct permutations is $\abs{C}!$. 

By making use of the above technical results, we are now in a position to show that every rate tuple $(\rK,r_V)$ in~\eqref{eq:MCH:R} is indeed achievable. That is, for any given $\rK$ in~\eqref{eq:MCH:R}, there is a discussion scheme for every $r_V$ in~\eqref{eq:MCH:R} that generates a secret key of rate $\rK$.

%\begin{Proof}[Theorem~\ref{thm:MCH:Region}]
Towards this goal, fix an arbitrary~$\rK$~in~\eqref{eq:MCH:R} and let $\mathscr{R}_V$ denote the set of all $r_V$ satisfying the inequalities in~\eqref{eq:MCH:R}. 
It follows that 
\begin{align}
\label{eq:Rv:Rc}
r_V\in\mathscr{R}_V\kern1em \text{ iff }\kern1em r_C\in \mathscr{R}_C,\forall C\in\mcP^*(\mcH).
\end{align} 
Let $q=|\mcP^*(\mcH)|$ and $\mcP^*(\mcH) = \left\{C_1, \ldots, C_q\right\}$. Then, consider any $r_V=\left(r_{C_1},\dots, r_{C_q}\right)\in \mathscr{R}_V$, by~\eqref{eq:Rv:Rc} and Corollary~\ref{lem:extreme_point}~\ref{pp:Pc:internal_point},  we have
\begin{align*}
r_{C_i}\geq \sum_{j}\alpha_{i j}\hat r_{C_i}^{(j)},\quad \forall C_i\in\mcP^*(\mcH),
\end{align*}
where $\alpha_{i j}\geq 0, \forall i,j,$ and $\sum_{j}\alpha_{i j}=1,\forall i$, and 
$\hat r_{C_i}^{(j)}$ is an extreme point of $\mathscr{R}_{C_i}$. 
It then follows that 
\begin{align*}
r_V\geq& \left(\sum_{j_1}\alpha_{1 j_1}\hat r_{C_1}^{(j_1)}, \dots,\sum_{j_q}\alpha_{q j_q}\hat r_{C_q}^{(j_q)} \right)\\
=& \sum_{j_1}\alpha_{1 j_1}\left(\hat r_{C_1}^{(j_1)}, \sum_{j_2}\alpha_{2 j_2}\hat r_{C_2}^{(j_2)},\dots,\sum_{j_q}\alpha_{q j_q}\hat r_{C_q}^{(j_q)} \right)\\
\vdots&\\
=&\sum_{j_1,\dots, j_q}\alpha_{1 j_1}\dots\alpha_{q j_q}\left(\hat r_{C_1}^{(j_1)},\dots,\hat r_{C_q}^{(j_q)}\right).
\end{align*}
Upon observing  
\begin{align*}
\sum_{j_1,\dots, j_q}\alpha_{1 j_1}\dots\alpha_{q j_q}=1,
\end{align*}
we conclude that $r_V$ is dominated by some convex combination of points $\left(\hat r_{C_1},\dots,\hat r_{C_q}\right)\in\mathscr{R}_V$, where each $\hat r_{C_i}$ is an extreme point of $\mathscr{R}_{C_i}$. (The number of such points is possibly $\prod_{i=1}^{q}\left(\abs{C_{i}}!\right)$.) To show there is a discussion scheme for $r_V$ that generates a secret key of rate $\rK$, it suffices to show there is a discussion scheme for every $\left(\hat r_{C_1},\dots,\hat r_{C_q}\right)$ that generates a secret key of rate $\rK$, because the usual time-sharing argument will then extend the schemes to the desired scheme for $r_V$. Further, it also suffices to show there is a discussion scheme for one such $\left(\hat r_{C_1},\dots,\hat r_{C_q}\right)$, because all others correspond to permutations of vertices, as substantiated by Corollary~\ref{lem:extreme_point}~\ref{pp:Pc:extreme_point}, and can therefore  be proved in the same manner. In what follows, we will give a discussion scheme for one $\left(\hat r_{C_1},\dots,\hat r_{C_q}\right)$ that enables the users to generate a secret key of rate $\rK$.

First, process each edge random variable  $\RX_{\rm{e}_i}$ such that 
\begin{align}
\label{eq:reduce_weight}
H(\tilde\RX_{\rm{e}_{i}})=\rK,\quad\forall i\in\Set{1,\dots,|E(\mcH)|},
\end{align}
where $\tilde\RX_{\rm{e}_{i}}$ denotes the uniformly distributed edge random variable after processing. This is possible as $$\rK\leq \CS(\infty)=\min_{\rm{e}\in E(\mcH)}H(\RX_{\rm{e}})$$ by~\eqref{eq:hyper:weight} and~\eqref{eq:MCH:CS}. 
Then, consider an arbitrary $C\in \mcP^*(\mcH)$. For notational simplicity, we assume
that $C=\left\{1,\dots,|C|\right\}$. For $\hat{r}_C=\left(\hat{r}_1,\dots,\hat{r}_{\abs{C}}\right)$ being an extreme point of $\mathscr{R}_C$, by virtue of the assertion~\ref{pp:Pc:extreme_point} of  Corollary~\ref{lem:extreme_point}, we can write 
\begin{subequations}
	\label{eq:one_extrem_point}
\begin{align}
\hat{r}_{1}&=[\kappa(\mcH/[1])-1]\rK,\\
\hat{r}_{i}&=[\kappa(\mcH/[i])-\kappa(\mcH/[i-1])]\rK, 
\end{align}
\end{subequations}
for $i=2,\dots, |C|$. 

To describe the discussion scheme for such $\hat{r}_C$, let us first introduce a few notations here. To proceed, recall the definition of $\mcH_{E_C}$~in~\eqref{def:eq:H_Ec}. 
Without loss of generality, let 
\begin{align}
	\label{eq:Ic}
\mathcal{I}_C=\Set{v_1,\dots, v_\ell}\kern0.5em \text{where} \kern0.5em
\ell=\kappa(\mcH_{E_C}/ C)
\end{align} 
be the set of representatives of the connected components in $\mcH_{E_C}/ C$. It follows from Lemma~\ref{lem:H_Ec}~\ref{H_Ec:i} that 
\begin{align}
	\label{eq:Ic:d}
\mathtt{d}_{\mcH_{E_C}}(v_s)=1,\quad\forall v_s\in \mathcal{I}_C.
\end{align} 
For $i\in C$, let 
\begin{align}
	\label{eq:Ic:i}
\mathcal{I}_C^i=\left\{\left. v_s\in\mathcal{I}_C\right|\exists \rm{e}\in E_C: \Set{i,v_s}\subseteq\xi_{\mcH_{E_C}}(\rm{e})\right\}
\end{align} 
be the subset of representatives that share an edge with vertex $i$ in $\mcH_{E_C}$. 
Note that $\mathcal{I}_C^i\neq \emptyset$ for all $i\in C$ by Lemma~\ref{lem:H_Ec}.
For $B\subseteq C$ and $v_s, v_t\in\mathcal{I}_C$, we will write $v_s\sim_{\mcH_{E_C}/B} v_t$ to indicate $v_t$ is reachable from $v_s$ via a path in hypergraph $\mcH_{E_C}/B$.
It follows that $\sim_{\mcH_{E_C}/[i]}$ is an equivalence relation and the set of equivalence classes of $\mathcal{I}_C^i$ induced is 
\begin{align*}
\mcP(\mathcal{I}_C^i)=\op{maximal}\left\{\left. S\subseteq \mathcal{I}_C^i\right| v_s\sim_{\mcH_{E_C}/[i]} v_t, \forall v_s,v_t\in S\right\}.%\in\Pi(\mathcal{I}_C^i)
\end{align*}
This is essentially the connected components of $\mcH_{E_C}/[i]$ but restricted to vertices in $\mathcal{I}_C^i$.

Now, we are ready to give the discussion scheme for $\hat{r}_C$ in~\eqref{eq:one_extrem_point}.   
Consider removing the vertices in $C$ successively in ascending order of index from $\mcH_{E_C}$. 
At the $i^{th}$ iteration, the discussion by user $i$ is as follows:% consider the following two cases:

\underline{Case 1:} $|\mcP(\mathcal{I}_C^i)|=1$, i.e., vertices in $\mathcal{I}_C^i$ still remain connected in $\mcH_{E_C}/ [i]$. In such a case, user $i$ should not discuss.

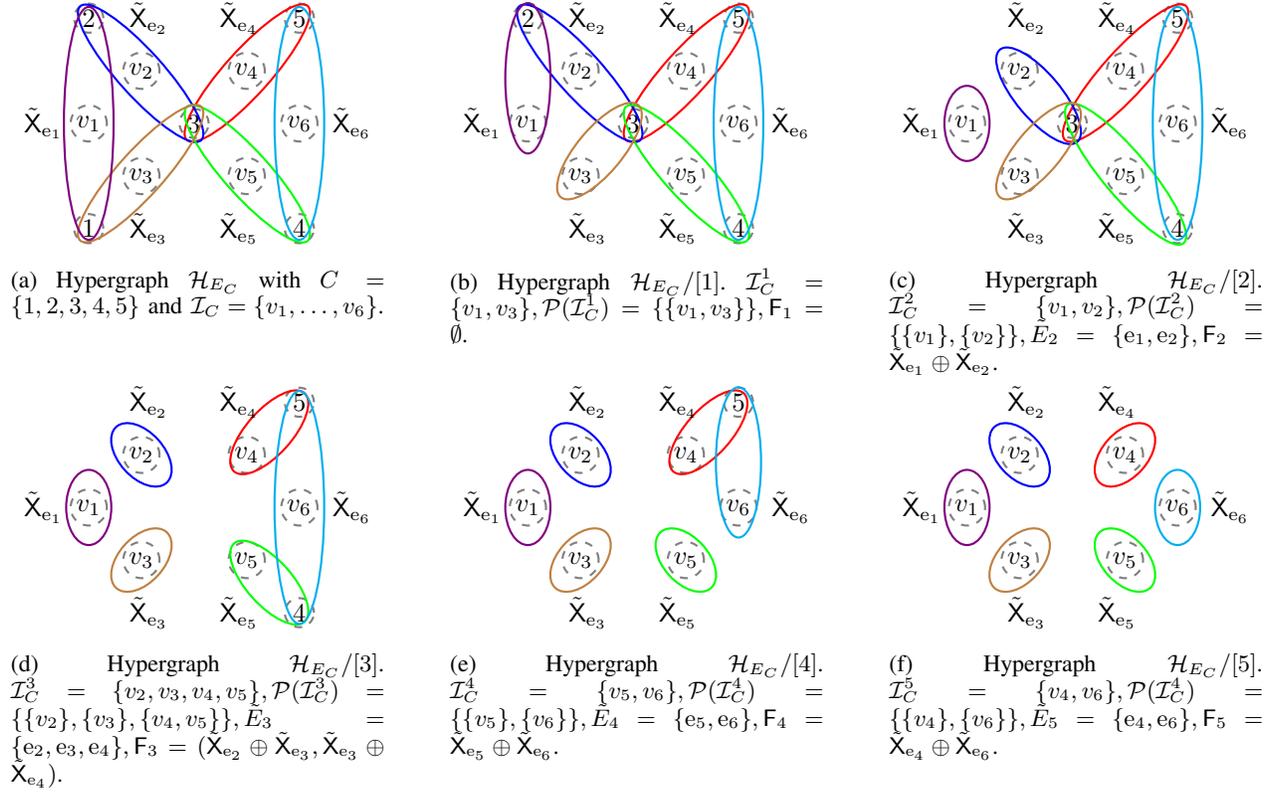
\begin{figure*}
	\centering
	\tikzstyle{dot}=[shape=circle,dashed,draw=gray!100,thick,inner sep=1pt,minimum size=11pt]
	\subcaptionbox{Hypergraph $\mcH_{E_C}$ with $C=\Set{1,2,3,4,5}$ and $\mathcal{I}_C=\Set{v_1,\dots, v_6}$.\label{fig:proof:0}}%[.3\textwidth]
	{\begin{tikzpicture}[scale=1]
	\node[dot] (3) at (0,0)  {$3$};
	\node[dot] (2) at (-1.4,1.4) {$2$};
	\node[dot] (5) at (1.4,1.4)  {$5$};
	\node[dot] (1) at (-1.4,-1.4) {$1$};
	\node[dot] (4) at (1.4,-1.4)  {$4$};
	\node[dot] (6) at (-.7,.7) {$v_2$};
	\node[dot] (7) at (.7,.7)  {$v_4$};
	\node[dot] (8) at (-.7,-.7) {$v_3$};
	\node[dot] (9) at (.7,-.7)  {$v_5$};
	\node[dot] (10) at (1.4,0)  {$v_6$};
	\node[dot] (11) at (-1.4,0)  {$v_1$};
   	\draw [thick,red,rotate around={48:(.6,0.7)}](.65,.6) ellipse (1.2cm and 0.3cm);
    	\draw [thick,blue,rotate around={132:(-.65,0.6)}](-.56,.6) ellipse (1.2cm and 0.3cm);
	\draw [thick,green,rotate around={-48:(.6,-0.7)}](.65,-.6) ellipse (1.2cm and 0.3cm);
	\draw [thick,brown,rotate around={-132:(-.65,-0.6)}](-.56,-.6) ellipse (1.2cm and 0.3cm);
        \draw [thick,violet,rotate around={90:(-1.4,0)}](-1.4,0) ellipse (1.55cm and 0.33cm);
        \draw [thick,cyan,rotate around={90:(1.4,0)}](1.4,0) ellipse (1.55cm and 0.33cm);
	\node[] () at (-2,0)  {$\tilde\RX_{\rm{e}_1}$};
	\node[] () at (-.6,-1.4) {$\tilde\RX_{\rm{e}_3}$};
	\node[] () at (-.6,1.4) {$\tilde\RX_{\rm{e}_2}$};
	\node[] () at (.6,-1.4) {$\tilde\RX_{\rm{e}_5}$};
	\node[] () at (.6,1.4) {$\tilde\RX_{\rm{e}_4}$};
	\node[] () at (2.1,0)  {$\tilde\RX_{\rm{e}_6}$};
	\end{tikzpicture}
	}
	\hfil
        \subcaptionbox{Hypergraph $\mcH_{E_C}/[1]$. $\mathcal{I}_C^1=\Set{v_1,v_3},  \mcP(\mathcal{I}_C^1)=\Set{\Set{v_1,v_3}}, 
        \RF_1=\emptyset$.\label{fig:proof:1}}
        {\begin{tikzpicture}[scale=1]
	\node[dot] (3) at (0,0)  {$3$};
	\node[dot] (2) at (-1.4,1.4) {$2$};
	\node[dot] (5) at (1.4,1.4)  {$5$};
	%\node[dot] (1) at (-1.4,-1.4) {$1$};
	\node[dot] (4) at (1.4,-1.4)  {$4$};
	\node[dot] (6) at (-.7,.7) {$v_2$};
	\node[dot] (7) at (.7,.7)  {$v_4$};
	\node[dot] (8) at (-.7,-.7) {$v_3$};
	\node[dot] (9) at (.7,-.7)  {$v_5$};
	\node[dot] (10) at (1.4,0)  {$v_6$};
	\node[dot] (11) at (-1.4,0)  {$v_1$};
   	\draw [thick,red,rotate around={48:(.6,0.7)}](.65,.6) ellipse (1.2cm and 0.3cm);
    	\draw [thick,blue,rotate around={132:(-.65,0.6)}](-.56,.6) ellipse (1.2cm and 0.3cm);
	\draw [thick,green,rotate around={-48:(.6,-0.7)}](.65,-.6) ellipse (1.2cm and 0.3cm);
	\draw [thick,brown,rotate around={-130:(-.3,-0.44)}](-.26,-.6) ellipse (.8cm and 0.3cm);
        \draw [thick,violet,rotate around={90:(-1.4,0.6)}](-1.4,.6) ellipse (1cm and 0.3cm);
        \draw [thick,cyan,rotate around={90:(1.4,0)}](1.4,0) ellipse (1.55cm and 0.33cm);
	\node[] () at (-2,0)  {$\tilde\RX_{\rm{e}_1}$};
	\node[] () at (-.6,-1.4) {$\tilde\RX_{\rm{e}_3}$};
	\node[] () at (-.6,1.4) {$\tilde\RX_{\rm{e}_2}$};
	\node[] () at (.6,-1.4) {$\tilde\RX_{\rm{e}_5}$};
	\node[] () at (.6,1.4) {$\tilde\RX_{\rm{e}_4}$};
	\node[] () at (2.1,0)  {$\tilde\RX_{\rm{e}_6}$};
	\end{tikzpicture}
	}
	\hfil
	\subcaptionbox{Hypergraph $\mcH_{E_C}/[2]$. $\mathcal{I}_C^2=\Set{v_1,v_2}, \mcP(\mathcal{I}_C^2)=\Set{\Set{v_1},\Set{v_2}}, \tilde{E}_2=\{\rm{e}_1,\rm{e}_2\},\RF_2=\tilde\RX_{\rm{e}_1}\oplus\tilde\RX_{\rm{e}_2}$.\label{fig:proof:2}}
	{\begin{tikzpicture}[scale=1]
	\node[dot] (3) at (0,0)  {$3$};
	%\node[dot] (2) at (-1.4,1.4) {$2$};
	\node[dot] (5) at (1.4,1.4)  {$5$};
	%\node[dot] (1) at (-1.4,-1.4) {$1$};
	\node[dot] (4) at (1.4,-1.4)  {$4$};
	\node[dot] (6) at (-.7,.7) {$v_2$};
	\node[dot] (7) at (.7,.7)  {$v_4$};
	\node[dot] (8) at (-.7,-.7) {$v_3$};
	\node[dot] (9) at (.7,-.7)  {$v_5$};
	\node[dot] (10) at (1.4,0)  {$v_6$};
	\node[dot] (11) at (-1.4,0)  {$v_1$};
   	\draw [thick,red,rotate around={48:(.6,0.7)}](.65,.6) ellipse (1.2cm and 0.3cm);
    	\draw [thick,blue,rotate around={130:(-.3,0.44)}](-.26,.6) ellipse (.8cm and 0.3cm);
	\draw [thick,green,rotate around={-50:(.6,-0.7)}](.65,-.6) ellipse (1.2cm and 0.3cm);
	\draw [thick,brown,rotate around={-130:(-.3,-0.44)}](-.26,-.6) ellipse (.8cm and 0.3cm);
        \draw [thick,violet,rotate around={90:(-1.4,0)}](-1.4,0) ellipse (.5cm and 0.3cm);
        \draw [thick,cyan,rotate around={90:(1.4,0)}](1.4,0) ellipse (1.55cm and 0.33cm);
	\node[] () at (-2,0)  {$\tilde\RX_{\rm{e}_1}$};
	\node[] () at (-.6,-1.4) {$\tilde\RX_{\rm{e}_3}$};
	\node[] () at (-.6,1.4) {$\tilde\RX_{\rm{e}_2}$};
	\node[] () at (.6,-1.4) {$\tilde\RX_{\rm{e}_5}$};
	\node[] () at (.6,1.4) {$\tilde\RX_{\rm{e}_4}$};
	\node[] () at (2.1,0)  {$\tilde\RX_{\rm{e}_6}$};
	\end{tikzpicture}
	}
\hfil
	\subcaptionbox{Hypergraph $\mcH_{E_C}/[3]$.
	 $\mathcal{I}_C^3=\Set{v_2,v_3,v_4,v_5}, 
	 \mcP(\mathcal{I}_C^3)=\Set{\Set{v_2},\Set{v_3},\Set{v_4,v_5}}, \tilde{E}_3=\{\rm{e}_2,\rm{e}_3,\rm{e}_4\},
	 \RF_3=(\tilde\RX_{\rm{e}_2}\oplus\tilde\RX_{\rm{e}_3},\tilde\RX_{\rm{e}_3}\oplus\tilde\RX_{\rm{e}_4})$.
	 \label{fig:proof:3}}
	{\begin{tikzpicture}[]
	%\node[dot] (3) at (0,0)  {$3$};
	%\node[dot] (2) at (-1.4,1.4) {$2$};
	\node[dot] (5) at (1.4,1.4)  {$5$};
	%\node[dot] (1) at (-1.4,-1.4) {$1$};
	\node[dot] (4) at (1.4,-1.4)  {$4$};
	\node[dot] (6) at (-.7,.7) {$v_2$};
	\node[dot] (7) at (.7,.7)  {$v_4$};
	\node[dot] (8) at (-.7,-.7) {$v_3$};
	\node[dot] (9) at (.7,-.7)  {$v_5$};
	\node[dot] (10) at (1.4,0)  {$v_6$};
	\node[dot] (11) at (-1.4,0)  {$v_1$};
   	\draw [thick,red,rotate around={48:(1,1)}](1,1) ellipse (.7cm and 0.3cm);
    	\draw [thick,blue,rotate around={132:(-.7,.7)}](-.7,.7) ellipse (.5cm and 0.3cm);
	\draw [thick,green,rotate around={-48:(1,-1)}](1,-1) ellipse (.7cm and 0.3cm);
	\draw [thick,brown,rotate around={-132:(-.7,-.7)}](-.7,-.7) ellipse (.5cm and 0.3cm);
        \draw [thick,violet,rotate around={90:(-1.4,0)}](-1.4,0) ellipse (.5cm and 0.3cm);
        \draw [thick,cyan,rotate around={90:(1.4,0)}](1.4,0) ellipse (1.55cm and 0.33cm);
	\node[] () at (-2,0)  {$\tilde\RX_{\rm{e}_1}$};
	\node[] () at (-.6,-1.4) {$\tilde\RX_{\rm{e}_3}$};
	\node[] () at (-.6,1.4) {$\tilde\RX_{\rm{e}_2}$};
	\node[] () at (.6,-1.4) {$\tilde\RX_{\rm{e}_5}$};
	\node[] () at (.6,1.4) {$\tilde\RX_{\rm{e}_4}$};
	\node[] () at (2.1,0)  {$\tilde\RX_{\rm{e}_6}$};
	\end{tikzpicture}
	}
\hfil
        \subcaptionbox{Hypergraph $\mcH_{E_C}/[4]$. 
        $\mathcal{I}_C^4=\Set{v_5,v_6}, \mcP(\mathcal{I}_C^4)=\Set{\Set{v_5},\Set{v_6}},
        \tilde{E}_4=\{\rm{e}_5,\rm{e}_6\}, \RF_4=\tilde\RX_{\rm{e}_5}\oplus\tilde\RX_{\rm{e}_6}$.\label{fig:proof:4}}
        {\begin{tikzpicture}[]
	%\node[dot] (3) at (0,0)  {$3$};
	%\node[dot] (2) at (-1.4,1.4) {$2$};
	\node[dot] (5) at (1.4,1.4)  {$5$};
	%\node[dot] (1) at (-1.4,-1.4) {$1$};
	%\node[dot] (4) at (1.4,-1.4)  {$4$};
	\node[dot] (6) at (-.7,.7) {$v_2$};
	\node[dot] (7) at (.7,.7)  {$v_4$};
	\node[dot] (8) at (-.7,-.7) {$v_3$};
	\node[dot] (9) at (.7,-.7)  {$v_5$};
	\node[dot] (10) at (1.4,0)  {$v_6$};
	\node[dot] (11) at (-1.4,0)  {$v_1$};
   	\draw [thick,red,rotate around={48:(1,1)}](1,1) ellipse (.7cm and 0.3cm);
    	\draw [thick,blue,rotate around={132:(-.7,.7)}](-.7,.7) ellipse (.5cm and 0.3cm);
	\draw [thick,green,rotate around={-48:(.7,-.7)}](.7,-.7) ellipse (.5cm and 0.3cm);
	\draw [thick,brown,rotate around={-132:(-.7,-.7)}](-.7,-.7) ellipse (.5cm and 0.3cm);
        \draw [thick,violet,rotate around={90:(-1.4,0)}](-1.4,0) ellipse (.5cm and 0.3cm);
        \draw [thick,cyan,rotate around={90:(1.4,0.6)}](1.4,0.6) ellipse (1cm and 0.3cm);
	\node[] () at (-2,0)  {$\tilde\RX_{\rm{e}_1}$};
	\node[] () at (-.6,-1.4) {$\tilde\RX_{\rm{e}_3}$};
	\node[] () at (-.6,1.4) {$\tilde\RX_{\rm{e}_2}$};
	\node[] () at (.6,-1.4) {$\tilde\RX_{\rm{e}_5}$};
	\node[] () at (.6,1.4) {$\tilde\RX_{\rm{e}_4}$};
	\node[] () at (2.1,0)  {$\tilde\RX_{\rm{e}_6}$};
	\end{tikzpicture}
	}
\hfil
	\subcaptionbox{Hypergraph $\mcH_{E_C}/[5]$. $\mathcal{I}_C^5=\Set{v_4,v_6},
	\mcP(\mathcal{I}_C^4)=\Set{\Set{v_4},\Set{v_6}},
	 \tilde{E}_5=\{\rm{e}_4,\rm{e}_6\}, \RF_5=\tilde\RX_{\rm{e}_4}\oplus\tilde\RX_{\rm{e}_6}$.\label{fig:proof:5}}
	{\begin{tikzpicture}[]
	%\node[dot] (3) at (0,0)  {$3$};
	%\node[dot] (2) at (-1.4,1.4) {$2$};
	%\node[dot] (5) at (1.4,1.4)  {$5$};
	%\node[dot] (1) at (-1.4,-1.4) {$1$};
	%\node[dot] (4) at (1.4,-1.4)  {$4$};
	\node[dot] (6) at (-.7,.7) {$v_2$};
	\node[dot] (7) at (.7,.7)  {$v_4$};
	\node[dot] (8) at (-.7,-.7) {$v_3$};
	\node[dot] (9) at (.7,-.7)  {$v_5$};
	\node[dot] (10) at (1.4,0)  {$v_6$};
	\node[dot] (11) at (-1.4,0)  {$v_1$};
   	\draw [thick,red,rotate around={48:(.7,.7)}](.7,.7) ellipse (.5cm and 0.3cm);
    	\draw [thick,blue,rotate around={132:(-.7,.7)}](-.7,.7) ellipse (.5cm and 0.3cm);
	\draw [thick,green,rotate around={-48:(.7,-.7)}](.7,-.7) ellipse (.5cm and 0.3cm);
	\draw [thick,brown,rotate around={-132:(-.7,-.7)}](-.7,-.7) ellipse (.5cm and 0.3cm);
        \draw [thick,violet,rotate around={90:(-1.4,0)}](-1.4,0) ellipse (.5cm and 0.3cm);
        \draw [thick,cyan,rotate around={90:(1.4,0)}](1.4,0) ellipse (.5cm and 0.3cm);
	\node[] () at (-2,0)  {$\tilde\RX_{\rm{e}_1}$};
	\node[] () at (-.6,-1.4) {$\tilde\RX_{\rm{e}_3}$};
	\node[] () at (-.6,1.4) {$\tilde\RX_{\rm{e}_2}$};
	\node[] () at (.6,-1.4) {$\tilde\RX_{\rm{e}_5}$};
	\node[] () at (.6,1.4) {$\tilde\RX_{\rm{e}_4}$};
	\node[] () at (2.1,0)  {$\tilde\RX_{\rm{e}_6}$};
	\end{tikzpicture}
	}
\caption{An example that illustrates our XOR discussion scheme for each $C\in \mcP^*(\mcH)$.}
\label{fig:proof:illus}
\end{figure*}

\underline{Case 2:} $|\mcP(\mathcal{I}_C^i)|>1$. In this case, for each element $S$ of $\mcP(\mathcal{I}_C^i)$, we randomly pick a representative $v_s\in S$. By the definitions~\eqref{eq:Ic}~and~\eqref{eq:Ic:i}, each representative shares an edge with vertex $i$ and the edges for different representatives are distinct. Let $\tilde{E}_i$ be the set of edges that are shared by these picked representatives and vertex $i$, say, $\tilde{E}_i=\Set{\rm{e}_1,\dots,\rm{e}_\lambda}$ with $\lambda=|\mcP(\mathcal{I}_C^i)|>1$. Then, user $i$ use the following scheme to discuss in public
\begin{align}
\label{eq:Fi:xor-struc}
\RF_i=(\tilde\RX_{\rm{e}_1}\oplus\tilde\RX_{\rm{e}_2},\dots,\tilde\RX_{\rm{e}_{\lambda-1}}\oplus\tilde\RX_{\rm{e}_{\lambda}})
\end{align}
Here, $\oplus$ refers to addition over corresponding finite field.
 %As an example,~\figref{fig:proof:illus} illustrates the above discussion scheme.
See~\figref{fig:proof:illus} for an illustration of the above discussion scheme.

Next, we show that the above discussion scheme has a discussion rate tuple $\hat{r}_C$ satisfying the rate constraint~\eqref{eq:one_extrem_point}.  
Observe that, at the $i^{th}$ iteration, the increase in the number of connected components after removing vertex $i$ from $\mcH_{E_C}/[i-1]$\footnote{For $i=1$, we use the convention that $[0]:=\emptyset$.} is completely determined by the connectedness of those vertices that share an edge with vertex $i$, which, by Lemma~\ref{lem:H_Ec}~\ref{H_Ec:ii}, can be
represented by the connectedness of vertices $\mathcal{I}_C^i$ in $\mcH_{E_C}/[i]$. 
It follows that
\begin{align}
\label{eq:F:Ei}
\kappa(\mcH_{E_C}/ [i])-\kappa(\mcH_{E_C}/ [i-1])=|\mcP(\mathcal{I}_C^i)|-1			
\end{align}
On the other hand, we have 
\begin{align}
\label{eq:k(H_Ec)=k(H)}
\kappa(\mcH/ [i])=\kappa(\mcH_{E_C}/ [i]), \quad\forall i\in C, 
\end{align}
which is by~\eqref{eq:cc:H_Ec/B=H/B} argued therein. 
Thus, on combining~\eqref{eq:Fi:xor-struc},~\eqref{eq:F:Ei} and~\eqref{eq:k(H_Ec)=k(H)}, we conclude that the above discussion scheme satisfies the rate constraint~\eqref{eq:one_extrem_point}.

Applying the above discussion scheme to all $C\in \mcP^*(\mcH)$ gives a discussion scheme for one $\left(\hat r_{C_1} ,\dots,\hat r_{C_q} \right)$. From~\eqref{eq:Fi:xor-struc},~\eqref{eq:F:Ei} and~\eqref{eq:k(H_Ec)=k(H)}, we know that the number of discussion by each $C$ is $\kappa(\mcH/ C)-1$. Therefore, the total number of discussion is 
\begin{align*}
	  \sum_{C\in \mcP^*(\mcH)}[\kappa(\mcH/ C)-1]
	  &=\sum_{C\in \mcP^*(\mcH)}[\mathtt{d}_{\mcH}(C)-1]\\
	  &=(\abs{E(\mcH)}-1)
\end{align*}
where the first equality follows from~\eqref{eq:mch:C(H)=d(C)}; and the last equality follows from~\eqref{eq:mch:d(C)=E}. Let $\mu=|E(\mcH)|$. It follows from~\eqref{eq:Fi:xor-struc} that the above discussion scheme can be expressed as
\begin{align}
\label{eq:F=Ax}
\RF=\MA\left[
	\begin{array}{c}
		\tilde\RX_{\rm{e}_{1}} \\	
		\vdots  \\	
		\tilde\RX_{\rm{e}_{\mu}}  \\	
	\end{array}
\right]
=\left[
	\begin{array}{c}
		\Ma_{1} \\	
		\vdots  \\	
		\Ma_{\mu-1}  \\	
	\end{array}
\right]\left[
	\begin{array}{c}
		\tilde\RX_{\rm{e}_{1}} \\	
		\vdots  \\	
		\tilde\RX_{\rm{e}_{\mu}}  \\	
	\end{array}
\right]
\end{align} 
where %$\tilde\RX_{e_{1}},\dots,\tilde\RX_{e_{|E(\mcH)|}}$ are the edge random variables after processing, $H(\tilde\RX_{e_{i}})=\rK,\forall i=1,\dots,|E(\mcH)|$, and 
$\MA$ is a matrix over $\Set{0,1}^{(\mu-1)\times \mu}$, each row $\Ma_i$ of $\MA$ has Hamming weight $2$. 

Now, it remains to show that the users can generate a secret key of rate $\rK$ from the discussion $\RF$ in~\eqref{eq:F=Ax}. To that end, we first argue that $\MA$ is full rank, i.e., $\rm{rank}(\MA)=\mu-1$.
To show $\rm{rank}(\MA)=\mu-1$, upon noting that each $\Ma_i$ in~\eqref{eq:F=Ax} has Hamming weight $2$, it suffices to assume each $\tilde\RX_{\rm{e}}$ is a Bernoulli $(\frac12)$ random variable and then show every discussion is independent of the remaining  discussions. More precisely, let 
\begin{align*}
\left[
	\begin{array}{c}
		\hat\RF_{1} \\	
		\vdots  \\	
		\hat\RF_{\mu-1}  \\	
	\end{array}
\right]=\MA\left[
	\begin{array}{c}
		\hat\RX_{\rm{e}_{1}} \\	
		\vdots  \\	
		\hat\RX_{\rm{e}_{\mu}}  \\	
	\end{array}
\right]
\end{align*}
where $\hat\RX_{\rm{e}_{i}}$'s are independent uniformly random bits,  we have $\rm{rank}(\MA)=\mu-1$ if $I(\hat\RF_{i}\wedge\hat\RF_{[\mu-1]\setminus\{i\}})=0,\forall i\in[\mu-1]$. Towards this end, consider the following two cases: 
%Note that $0,1$ is a subfield of any finite filed. therefore, the procedure is valid.  
\begin{itemize}
\item Independence inside each $C$:  Observe that, by successively removing the vertices in $C$ from $\mcH_{E_C}$, all $v_i$'s in $\mathcal{I}_C$ will eventually become disconnected. Therefore, according to the discussion scheme, the $\ell=\kappa(\mcH_{E_C}/ C)$ edges incident on $\mathcal{I}_C$ will be involved in the discussions. 
By~\eqref{eq:Fi:xor-struc} and~\eqref{eq:F:Ei}, there are in total $\kappa(\mcH_{E_C}/ C)-1$ discussions by users in $C$. Define a graph $\mcG$ as follows: View the $\ell$ hyperedges as vertices and draw an edge between two hyperedges $\rm{e}$ and $\rm{e}'$ if they are involved in a discussion, i.e., $\hat\RX_{\rm{e}}\oplus\hat\RX_{\rm{e}'}$. From Lemma~\ref{lem:H_Ec}~\ref{H_Ec:i}~and~\ref{H_Ec:ii}, we know that $\mcG$ is connected. Therefore, $\mcG$ is a tree since it connects $\ell$ vertices (hyperedges) with $\ell-1$ edges.
Now, suppose to the contrary that a discussion by a user in $C$, say, $\hat\RX_{\rm{e}}\oplus\hat\RX_{\rm{e}'}$, is correlated with the remaining discussions by some users in $C$. By linearity of the discussion, it means a sequence of discussions by users in $C$ will determine $\hat\RX_{\rm{e}}\oplus\hat\RX_{\rm{e}'}$, i.e., we have a telescoping sum
\begin{align*}
    (\hat\RX_{\rm{e}}\oplus\hat\RX_{\rm{e}'}) &= (\hat\RX_{\rm{e}}\oplus\hat\RX_{\rm{e}_1})\oplus (\hat\RX_{\rm{e}_1}\oplus\hat\RX_{\rm{e}_2})\oplus\cdots\oplus\\
    &\kern1em \oplus (\hat\RX_{\rm{e}_{i}}\oplus\hat\RX_{\rm{e}_{i+1}}) \oplus \cdots \oplus (\hat\RX_{\rm{e}_{m-1}}\oplus\hat\RX_{\rm{e'}}),
\end{align*}
where each XOR within a pair of parentheses is a discussion by a user in $C$, and
 $m$ denotes the number of discussions on the right by some users in $C$. We allow $m=1$, in which case the sum on the right is simply $\hat\RX_{e}\oplus\hat\RX_{e'}$, i.e., there are two users in $C$ repeating the same message during the discussion. %They should be of the form\footnote{For $m=1$, it should be $\tilde\RX_{e}\oplus\tilde\RX_{e'}$.}
% \begin{align*}
% \hat\RX_{\rm{e}}\oplus\hat\RX_{\rm{e}_1},\dots,\hat\RX_{\rm{e}_i}\oplus\hat\RX_{\rm{e}_j},\dots,\hat\RX_{\rm{e}_{m-1}}\oplus\hat\RX_{\rm{e}'}, 
% \end{align*}
% where 
%\begin{itemize}[leftmargin=*]%[label=\roman*)]
Without loss of generality, we can assume that
\begin{itemize}
\setlength{\itemindent}{.1in}
\item the two edges involved in each XOR are distinct;  
\item the number of times each $\hat\RX_{\rm{e}_i}$ appeared is even.
\end{itemize}
%Then, the edges involved in the $\tilde\RX_{\rm{e}}\oplus\tilde\RX_{\rm{e}'}$ XOR discussion and the $m$ XOR discussions should form at least one cycle in $\mcH_{E_C}$, which involves $\rm{e}$ and $\rm{e}'$. 
It follows that $(\rm{e},\rm{e}_1), (\rm{e_1},\rm{e}_2),\dots, (\rm{e}_{m-1},\rm{e}')$ together with $(\rm{e},\rm{e}')$ forms at least one cycle in $\mcG$.
However, this contradicts $\mcG$ is a tree.  Thus, the discussions by $C$ are independent of each other. 
\item Independence between different $C$: Since $\mcH[\mcP^*(\mcH)]$ is a hypertree by Lemma~\ref{lem:MCH}~\ref{mch-to-hypertree}, we can use the reordering method as in the proof of Lemma~\ref{lem:HT} to enumerate $\mcP^*(\mcH)$ such that $C_{i+1}$ and $\bigcup_{j=1}^i C_j$ share exactly one edge for all $1\leq i\leq q-1$, i.e., satisfying~\eqref{eq:C:e:1}. It follows that
\begin{align*}
H(\hat\RF)=&H(\hat\RF_{C_1}, \dots,\hat\RF_{C_q})\\
	  =&H(\hat\RF_{C_q})+H(\hat\RF_{C_1}, \dots,\hat\RF_{C_{q-1}}|\hat\RF_{C_q})\\
	  =&H(\hat\RF_{C_q})+H(\hat\RF_{C_1}, \dots,\hat\RF_{C_{q-1}})\\
	  \vdots&\\
	 =&\sum_{i=1}^q H(\hat\RF_{C_i})
\end{align*} 
 i.e., the discussions by different $C$ are also independent of each other. 
\end{itemize}
Summarizing the above two cases, we conclude that every discussion is independent of the remaining discussions, thereby $\rm{rank}(\MA)=\mu-1$.

With this, we proceed to show that the users can generate a secret key of rate $\rK$ based on the discussion $\RF$ in~\eqref{eq:F=Ax}.
For $i\in[\mu]$, let $\mathbf{b}_i=(b_{i1},\dots,b_{i\mu})\in\Set{0,1}^{\mu}$ be an indicator vector, where $b_{ii}=1$ and $b_{ij}=0,j\neq i$. 
It then follows that 
\begin{align*}
\bM \MA \\ \mathbf{b}_i \eM \left[
	\begin{array}{c}
		\tilde\RX_{\rm{e}_{1}} \\	
		\vdots  \\	
		\tilde\RX_{\rm{e}_\mu}  \\	
	\end{array}
\right]=\bM \RF \\  \tilde\RX_{\rm{e}_{i}}\eM 
\end{align*} 
Next, we will argue that $\rm{rank}\left(\bM \MA \\ \mathbf{b}_i \eM\right)=\mu, \,\forall i\in[\mu]$, i.e., full rank. 
Suppose to the contrary that $\bM \MA \\ \mathbf{b}_i \eM$ is not full rank. Since $\rm{rank}(\MA)=\mu-1$, $\mathbf{b}_i$ must be in the row span of $\MA$ and so 
\begin{align*}
\exists S\subseteq[\mu-1], \quad \mathbf{b}_i =\sum_{i\in S}\Ma_{i}, 
\end{align*}
which gives
\begin{align}
\label{eq:indep:contra}
\sum_{j=1}^{\mu}b_{ij}&=\sum_{i\in S}\sum_{j=1}^{\mu}a_{ij}\notag\\
1=\sum_{j=1}^{\mu}b_{ij}&=\sum_{i\in S}\sum_{j=1}^{\mu}a_{ij}=0
\end{align}
yielding a contradiction. Here, the first equality of~\eqref{eq:indep:contra} follows from the definition of $\mathbf{b}_i $; the last equality of~\eqref{eq:indep:contra} is because the Hamming weight of $\Ma_{i}$ is $2$, $\forall i\in[\mu-1]$. Hence, $\bM \MA \\ \mathbf{b}_i \eM$ is full rank for all $i\in[\mu]$.
Since every user observes at least one edge random variable, say, $\tilde\RX_{\rm{e}_{i}}$, the discussion $\RF$ in~\eqref{eq:F=Ax} enables him to recover $[\tilde\RX_{\rm{e}_{1}},\dots,\tilde\RX_{\rm{e}_{\mu}}]^T$ as $\bM \MA \\ \mathbf{b}_i \eM^{-1}\bM \RF \\  \tilde\RX_{\rm{\rm{e}}_{i}}\eM$. It remains to extract a secret key of rate $\rK$ from the obtained $[\tilde\RX_{\rm{e}_{1}},\dots,\tilde\RX_{\rm{e}_{\mu}}]^T$. Towards this end, upon noting $\rm{rank}(\MA)=\mu-1$, by~\eqref{eq:F=Ax}, we have  
\begin{align*}
H(\RF)=(\mu-1)\rK.
\end{align*}
It then follows that
\begin{align*}
I(\tilde\RX_{\rm{e}_{i}}\wedge\RF)&=H(\tilde\RX_{\rm{e}_{i}})+H(\RF)-H(\tilde\RX_{\rm{e}_{i}},\RF)\\
					    &=H(\tilde\RX_{\rm{e}_{i}})+H(\RF)-H(\tilde\RX_{\rm{e}_{1}},\dots,\tilde\RX_{\rm{e}_{\mu}})\\
					    &=0, \kern3em\forall i=1,\dots,\mu
\end{align*} 
where the second equality follows from the recoverability argued above; the last equality follows from~\eqref{eq:reduce_weight}. Thus, we can choose any $\tilde\RX_{\rm{e}_{i}}$ as the secret key since it satisfies the perfect secrecy condition~\eqref{eq:recover}, \eqref{eq:secrecy} and \eqref{eq:perfect}. 
%Therefore, by~\eqref{eq:reduce_weight}, a secret key of rate $\rK$, 
%\begin{enumerate*}[label=(\roman*)]
%  \item has been attained zero error, and perfect secrecy through the above linear non-interactive discussion scheme;
%  \item has been attained non-asymptotically with block length $n=1$ and chosen to be function of any edge variable $\RX_e, e\in E(\mcH)$. 
%\end{enumerate*}
Therefore, by~\eqref{eq:reduce_weight}, a secret key of rate $\rK$, which can be chosen to be a function of an arbitrary edge variable $\RX_\rm{e}, \rm{e}\in E(\mcH)$, has been attained non-asymptotically with block length $n=1$, zero error, and perfect secrecy through the above linear non-interactive discussion scheme.
This proves the achievability of~\eqref{eq:MCH:R}, and the assertions in Theorem~\ref{thm:MCH:Region}. 
%\end{Proof}

\subsection{Proof of Corollary~\ref{cor:HT}}
%\begin{Proof}[Corollary~\ref{cor:HT}]
For $\mcH$ being a hypertree, we have $$\mcP^*(\mcH)=\Set{\Set{i}\mid i\in V}$$ by~\eqref{eq:HT:P*} in Lemma~\ref{lem:HT}. Then,~\eqref{eq:MCH:R} in~Theorem~\ref{thm:MCH:Region} will reduce to
\begin{align*}
r_i &\geq [\kappa(\mcH/\Set{i})-1)]\rK\\
     &=\left[\mathtt{d}_{\mcH}(i)-1\right]\rK 
\end{align*}
for $i\in V$, where the equality follows from~\eqref{eq:mch:C(H)=d(C)} in Lemma~\ref{lem:MCH}. This completes the proof of Corollary~\ref{cor:HT}.
%\end{Proof}

\subsection{Proof of Theorem~\ref{thm:MCH:CS(R)}}
%\begin{Proof}[Theorem~\ref{thm:MCH:CS(R)}]
Let $q=\abs{\mcP^{*}(\mcH)}$ and $\mcP^*(\mcH)=\{C_1,\dots,C_q\}$. Consider an arbitrary $\rK$ in~\eqref{eq:MCH:R}. We know from the achievability proof of Theorem~\ref{thm:MCH:Region} that any $r_V$ satisfying~\eqref{eq:MCH:R} is dominated by some convex combination of points $(\hat r_{C_1},\dots,\hat r_{C_q})$, where each $\hat r_{C_i}$ is an extreme point of $\mathscr{R}_{C_i}$. For each $C_i$, by the assertion~\ref{pp:Pc:extreme_point} of Corollary~\ref{lem:extreme_point}, it is easy to check that for each of those $|C_i|!$ extreme points of $\mathscr{R}_{C_i}$, the sum rate $$\hat r(C_i)=[\kappa(\mcH/ C_i)-1]\rK.$$ Therefore, the minimum sum rate required for generating a secret key of rate $\rK$ is   
\begin{align*}
       \sum_{C\in \mcP^*(\mcH)}\hat{r}(C) 
      &=\sum_{C\in \mcP^*(\mcH)}[\kappa(\mcH/ C)-1]\rK \\
      &=\sum_{C\in \mcP^*(\mcH)}[\mathtt{d}_{\mcH}(C)-1]\rK \\
      &=(\abs{E(\mcH)}-1)\rK
\end{align*}
where the second last equality follows from~\eqref{eq:mch:C(H)=d(C)}; and the last equality follows from~\eqref{eq:mch:d(C)=E}.
This, along with the fact that $\CS(R)\leq \CS(`8),\forall R\geq0$, yields~\eqref{eq:MCH:CS(R)}.
This completes the proof of Theorem~\ref{thm:MCH:CS(R)}.
%\end{Proof}

\section{Conclusion}
\label{sec:con}
We consider the problem of secret key generation in the multiterminal source model, subject to limited discussion rates. 
For sources that can be represented by minimally connected hypergraphs, a single-letter explicit characterization of the region of achievable secret key rate and public discussion rate tuple is established. Furthermore, we show that 
%the secret key can be attained with zero error and perfect secrecy through the above linear non-interactive discussion scheme, 
%and can be chosen to be function of any edge variable $\RX_e, e\in E(\mcH)$ . 
%\begin{enumerate*}%[label=(\roman*)]
the secret key can be attained with zero error and perfect secrecy through linear non-interactive discussion, without any additional cost in the discussion rates.
We also point out that the secret key can be attained non-asymptotically with only one observation, and chosen to be a function of an arbitrary edge random variable. 
%\end{enumerate*}
Finally, we obtained a closed-form formula for the maximum achievable secret key rate under any given total public discussion rate for such kind of sources. It turns out the optimal trade-off is characterized simply by the number of edges. 

In the course of deriving the main results, we clarify some combinatorial properties of hypergraphs. Specifically, we find a particular method that can reduce the minimally connected hypergraph to the hypertree. This reduction elucidates that the paths between certain sets of vertices are unique. As a consequence, the secret key can be optimally propagated along those paths. Besides, we also show that for certain subhypergraphs, the number of connected components is supermodular, while for some other parts, the normalization of the number of connected components by minus one  is subadditive. These combinatorial structures are established purely through the hypergraph notions and operations, and appear to be of fundamental interest in graph theory. 

However, for the hypergraphical sources where the corresponding hypergraph is not minimally connected, the techniques considered do not directly extend. One interesting direction is to characterize the achievable rate region of the graphical sources, i.e., the PIN model, for which the optimal trade-off between the achievable secret key rate and the total discussion rate has been resolved recently by~\cite{chan17isit,chan18LB}.
%Instead, diving into the hypergraph, it is worthwhile to consider graph first. 

\appendices
\makeatletter
\@addtoreset{equation}{section}
\@addtoreset{Theorem}{section}
\renewcommand{\theTheorem}{\thesection.\arabic{Theorem}}
\renewcommand{\theequation}{\thesection.\arabic{equation}}
\renewcommand{\theparentequation}{\thesection.\arabic{parentequation}}
\@addtoreset{Lemma}{section}
\renewcommand{\theLemma}{\thesection.\arabic{Lemma}}
\@addtoreset{Corollary}{section}
\renewcommand{\theCorollary}{\thesection.\arabic{Corollary}}
\@addtoreset{Example}{section}
\renewcommand{\theExample}{\thesection.\arabic{Example}}
\@addtoreset{Proposition}{section}
\renewcommand{\theProposition}{\thesection.\arabic{Proposition}}
\makeatother

\section{Proof of Lemma~\ref{lem:MCH}}
\label{sec:proof:lem:MCH}

The proof of Lemma~\ref{lem:MCH} takes recourse to the following technical result. %Lemma~\ref{lem:HT}. 
\begin{Lemma}
	\label{lem:HT}
A hypergraph $\mcH$ is connected and cycle-free iff 
\begin{subequations}
\label{eq:HT:I&P}
\begin{align}
I(\mcH)&=1,\kern1em \text{and} \label{eq:HT:I=1}\\ 
\mcP^*(\mcH)&=\Set{\Set{v}\mid v\in V(\mcH)}, \label{eq:HT:P*}
\end{align}
\end{subequations}
i.e., singleton partition is the fundamental partition.
\end{Lemma}
Lemma~\ref{lem:HT} provides an alternative characterization of the hypergraphs which admit unique path between every pair of distinct vertices through the notions of partition connectivity and fundamental partition. It is worth mentioning that a hypertree is a connected and cycle-free hypergraph but the reverse does not hold as the latter is allowed to contain loops. See the following simple example for illustration.
%\begin{Proof}
%See Appendix~\ref{sec:proof:hypergraph}.
%\end{Proof}
\begin{Example}
\label{eg:HT}
Consider the hypergraph in~\figref{fig:h:loop}. Compared to the hypertree in~\figref{fig:ht}, the difference is that there is a loop at vertex $3$ and vertex $5$, respectively. Observe that the path between any two distinct vertices is unique. It then follows from the above result that $I(\mcH)=1$ and $\mcP^*(\mcH)=\Set{\Set{1},\Set{2},\Set{3},\Set{4},\Set{5}}$. 
\end{Example}
\begin{figure}
\centering
	\tikzstyle{dot}=[shape=circle,dashed,draw=gray!100,thick,inner sep=1pt,minimum size=11pt]
\begin{tikzpicture}
	\node[dot] (1) at (-0.2,0) {$1$};
	\node[dot] [right= 0.8cm of 1] (2) {$2$};
	\node[dot] [right= 0.8cm of 2] (3) {$3$};
	\node[dot] [right= 0.8cm of 3] (4) {$4$};
	\node[dot] [right= -1.8cm of 1] (5) {$5$};
        \draw [thick,red](1.0,0)[] ellipse (1.8cm and 0.5cm);
        \draw [thick,blue](2.9,0)[] ellipse (1cm and 0.4cm);
        \draw [thick,violet](-0.9,0)[] ellipse (1cm and 0.4cm);
       	\draw [thick,green,rotate around={90:(-1.6,0.5)}](-1.6,0.5) ellipse (0.9cm and .5cm);
       	\draw [thick,brown,rotate around={90:(2.2,0.5)}](2.2,0.5) ellipse (0.9cm and .5cm);
	\node[] () at (1.0,-0.65)  {$\rm{a}$};		
	\node[] () at (3.0,-0.65)  {$\rm{b}$};		
	\node[] () at (-1.0,-0.65) {$\rm{c}$};		
	\node[] () at (1.6,0.9) {$\rm{d}$};		
	\node[] () at (-1.0,0.9) {$\rm{e}$};		
\end{tikzpicture}
\caption{A hypergraph $\mcH$ with $V(\mcH)=\Set{1,2,3,4,5}, E(\mcH)=\Set{\rm{a},\rm{b},\rm{c},\rm{d},\rm{e}}$ and $\xi_{\mcH}(\rm{a})=\Set{1,2,3},\xi_{\mcH}(\rm{b})=\Set{3,4},\xi_{\mcH}(\rm{c})=\Set{1,5},\xi_{\mcH}(\rm{d})=\Set{2},\xi_{\mcH}(\rm{e})=\Set{5}$.}
\label{fig:h:loop}
\end{figure}

\begin{Proof}[Lemma~\ref{lem:HT}]
We first prove the ``if" part. Consider a hypergraph $\mcH$ that satisfies~\eqref{eq:HT:I&P}. $I(\mcH)=1$ implies that $\mcH$ is connected. It remains to argue the path between any two distinct vertices in $\mcH$ is unique, i.e., no cycles. Suppose to the contrary that there exists a cycle in $\mcH$, say, $(v_1,\rm{e}_1,v_2,\dots,\rm{e}_{\ell-1},v_\ell=v_1)$ with $\ell\geq 3$. Define a graph $\mcG=(V,E,\xi)$ with 
\begin{align*}
V(\mcG)&=\Set{v_1,\dots, v_{\ell-1}}, \\
E(\mcG)&=\Set{\rm{e}_1,\dots,\rm{e}_{\ell-1}}, \\
\xi_\mcG(\rm{e}_i)&=\begin{cases}
\Set{v_{i},v_{i+1}},& i \in \Set{1,\dots,\ell-2},\\
\Set{v_{i},v_{1}},& i=\ell-1.
\end{cases}
\end{align*}
i.e., $\mcG$ is obtained from $\mcH_{\Set{v_1,\dots, v_{\ell-1}}}$ by further shrinking the hyperedges into edges and removing the hyperedges not in the sequence.
Then, we have
\begin{align}
	\label{eq:cycle:IH>1}
I\!\left(\mcH_{ \Set{v_1,\dots, v_{\ell-1}}}\right)\utag{a}\geq I(\mcG) 
						    \utag{b}>1
						    \utag{c}=I(\mcH)
\end{align}
where \uref{a} can be argued as follows. On one hand, shrinking the hyperedges into edges does not change the number of edges but will reduce the degree of certain nodes. Hence, it can only decrease the partition connectivity. On the other hand, from~\eqref{eq:I(H)}, it is obvious that removing the hyperedges can only decrease the partition connectivity. Altogether, we have \uref{a} as desired. 
\uref{b} is because $$I(\mcG)=\frac{\ell-1}{\ell-2}>1,$$ which is achieved by $\mcP^*(\mcG)=\Set{\Set{v_i}\mid i\in \Set{1,\dots,\ell-1}}$; \uref{c} is because $I(\mcH)=1$ by the assumption. Then, by Proposition~\ref{pro:P*}, we know that the fundamental partition is not the singleton partition, which contradicts our assumption~\eqref{eq:HT:P*}. 

We now prove the ``only if" part. Suppose $\mcH$ is connected and cycle-free. Consider an arbitrary $\mcP\in\Pi'(V)$, let $q=|\mcP|$. Since $\mcH$ is connected and cycle-free, we can enumerate $\mcP$ as $\{C_1,\dots,C_q\}$ such that $C_{i+1}$ and $\bigcup_{j=1}^i C_j$ share exactly one edge for all $1\leq i\leq q-1$, i.e.,
\begin{multline}
\label{eq:C:e:1}
\forall\, 1\leq i\leq q-1, \exists!\, \rm{e}\in E(\mcH) \\
\text{ s.t. }\xi_{\mcH}(\rm{e})\cap C_{i+1}\neq \emptyset, \xi_{\mcH}(\rm{e})\cap\bigcup_{j=1}^i C_j \neq \emptyset.
\end{multline} 
This can be done via reordering as follows:
Let $\mcP = \{C_1, \ldots, C_q\}$ be an arbitrary enumeration of the elements in the partition. We are going to construct a permutation $\pi: [q]\to [q]$ to reorder $\mcP$ such that it satisfies~\eqref{eq:C:e:1}. First, define $$\pi(1)=1.$$
Then, for $i$ from $2$ to $q$,  pick an element $C_\ell\in\mcP\setminus \{C_{\pi^{-1}(1)},\dots,C_{\pi^{-1}(i-1)}\}$ such that it shares exactly one edge with the set of all the previous picked elements $\bigcup_{j=1}^{i-1} C_{\pi^{-1}(j)}$. The existence of such an element $C_\ell$ is guaranteed by the connectedness of $\mcH$. The uniqueness of the edge between $C_\ell$ and $\bigcup_{j=1}^{i-1} C_{\pi^{-1}(j)}$ is guaranteed by the fact that $\mcH$ is cycle-free. Define $$\pi(\ell)=i.$$
It follows from the above construction that the reordered $C_{\pi^{-1}(1)},\dots,C_{\pi^{-1}(q)}$ satisfies the desired property~\eqref{eq:C:e:1}. 

Now, assume~\eqref{eq:C:e:1} holds. Upon using chain rule expansion, we can rewrite $E_{\mcP}(\mcH)$ as 
\begin{align*}
&\mkern5mu E_{\mcP}(\mcH)\\
&=\sum_{i=1}^{q-1}\left[\mathtt{d}_{\mcH}\Bigg(\bigcup_{j=1}^iC_j\Bigg)+\mathtt{d}_{\mcH}(C_{i+1})-\mathtt{d}_{\mcH}\Bigg(\bigcup_{j=1}^{i+1}C_j\Bigg)\right]\\
&=\sum_{i=1}^{q-1}\left|\mkern-3mu\Bigg\{\rm{e}\in E(\mcH)\mid \xi_{\mcH}(\rm{e})\cap\bigcup_{j=1}^i C_j \mkern-3mu\neq \mkern-3mu\emptyset,\xi_{\mcH}(\rm{e})\cap C_{i+1}\mkern-3mu\neq \mkern-3mu\emptyset \Bigg\}\mkern-3mu\right|\\
&=\sum_{i=1}^{q-1}1\\
&=q-1
\end{align*}
where the second last equality follows from~\eqref{eq:C:e:1}. 
Since the above holds for arbitrarily $\mcP$, we have~\eqref{eq:HT:I=1} as desired.  
Consider any $C\subseteq V(\mcH)$ with $\abs{C}>1$, $\mcH_{C}$ is a hypergraph with at most one path between any two distinct vertices since $\mcH$ is cycle-free, i.e., $\mcH_{C}$ is either connected and cycle-free or is disconnected. For $\mcH_{C}$ being connected and cycle-free, $I(\mcH_{C})=1$ as argued above. For $\mcH_{C}$ being disconnected, $I(\mcH_{C})=0$. Putting it all together, 
\begin{align*}
I(\mcH_{C})\leq1=I(\mcH), \quad\forall C\subseteq V(\mcH): |C|>1. 
\end{align*}
Then, by Proposition~\ref{pro:P*}, we have~\eqref{eq:HT:P*} as desired. This completes the proof of Lemma~\ref{lem:HT}.
\end{Proof}

Now, with Lemma~\ref{lem:HT} in hand, we proceed to the proof of Lemma~\ref{lem:MCH}.
%\begin{Proof}[Lemma~\ref{lem:MCH}]
We first prove the ``if" case of the first assertion. Suppose $\mcH[\mcP^*(\mcH)]$ is a hypertree. 
\begin{itemize}
\item Connectedness inside each $C\in \mcP^*(\mcH)$: By Proposition~\ref{pro:P*}, we have that 
\begin{align*}
I(\mcH_{C})> I(\mcH)\geq 0,\quad\forall C\in \mcP^*(\mcH): \abs{C}>1.
\end{align*} 
Therefore, $\mcH_{C}$ is connected. 
\item Connectedness between different $C\in \mcP^*(\mcH)$: Since $\mcH[\mcP^*(\mcH)]$ is 
a hypertree, vertices in $\mcP^*(\mcH)$ are connected.%, i.e., $\forall \mcS\subsetneq\mcP^*(\mcH)$, $\exists \rm{e}\in E(\mcH)$ such that $\xi_{\mcH}(e)\cap \bigcup_{C\in \mcS}C\neq \emptyset$ and $\xi_{\mcH}(e)\setminus \bigcup_{C\in \mcS}C\neq \emptyset$. 
\end{itemize}
Altogether, we have $\mcH$ is also connected. Since $\mcH[\mcP^*(\mcH)]$ is a hypertree and therefore loopless, every edge $\rm{e}\in E(\mcH[\mcP^*(\mcH)])$ is incident on at least two distinct vertices in $\mcP^*(\mcH)$, say, $C'$ and $C''$. Then, $\exists \,v'\in C'$ and $\exists v''\in C''$ such that $v',v''\in \xi_{\mcH}(\rm{e})$ by Definition~\ref{def:H[P]} of $\mcH[\mcP^*(\mcH)]$. It follows that $(v',\rm{e},v'')$ is a path for $v'$ and $v''$ in $\mcH$. It is unique because $(C',\rm{e},C'')$ is a unique path for $C'$ and $C''$ in the hypertree $\mcH[\mcP^*(\mcH)]$. Removing edge $\rm{e}$ from $\mcH$ will disconnect $v'$ and $v''$. Lastly, upon noting that $E(\mcH[\mcP^*(\mcH)])=E(\mcH)$, we conclude that $\mcH$ is minimally connected. 

Now we prove the ``only if" case of the first assertion. When $\mcH$ is minimally connected, then $\mcH[\mcP^*(\mcH)]$ is also connected by its Definition~\ref{def:H[P]}. Next, we proceed to prove that $\mcH[\mcP^*(\mcH)]$ is cycle-free. Let $\mcP$ be the set of equivalent classes (or connected components) of $\mcH$ after removing any edge $\rm{e}$. Since $\mcH$ is minimally connected, we have $\mcP\in \Pi'(V)$. Then, it follows that 
\begin{align}
\label{eq:I(H)<=1}
I(\mcH)\leq \frac1 {\abs{\mcP}-1}E_{\mcP}(\mcH)=1,
\end{align}
where the equality follows from the fact that only edge $\rm{e}$ crosses $\mcP$, i.e., $E_{\mcP}(\mcH)=\abs{\mcP}-1$. 
Now, suppose to the contrary that there is a cycle in $\mcH[\mcP^*(\mcH)]$, then there is also a cycle in $\mcH$, say, $(v_1,\rm{e}_1,v_2,\dots,\rm{e}_{\ell-1},v_{\ell}=v_1)$ with $\ell\geq3$, that crosses $\mcP^*(\mcH)$, i.e., $\Set{v_1,\dots,v_{\ell-1}}\not\subseteq C, \forall C\in \mcP^*(\mcH)$. Then, we have that
\begin{align*}
I(\mcH_{\Set{v_1,\dots, v_{\ell-1}}})\utag{a}>1 \utag{b}\geq I(\mcH) 
\end{align*}
where \uref{a} is by~\eqref{eq:cycle:IH>1} argued before; \uref{b} is by~\eqref{eq:I(H)<=1} argued above. However, this violates the Proposition~\ref{pro:P*}. Therefore, $\mcH[\mcP^*(\mcH)]$ is cycle-free. What remains to be shown is that $\mcH[\mcP^*(\mcH)]$ is loopless. Suppose to the contrary that $\mcH[\mcP^*(\mcH)]$ has a singleton edge $\rm{e}\in E(\mcH[\mcP^*(\mcH)])$ incident on a vertex $C\in \mcP^*(\mcH)$. By Definition~\ref{def:H[P]}, $\rm{e}\in E(\mcH)$.  
%It follows from Definition~\ref{def:H[P]} that $$\xi_{\mcH}(\rm{e})\subseteq \xi_{\mcH[\mcP^*(\mcH)]}(\rm{e})=C.$$ Hence, removing edge $\rm{e}$ from $\mcH$ can only affect the connectedness 
Removing edge $\rm{e}$ from $\mcH$, we have 
\begin{itemize}
\item Connectedness inside $C$: $\mcH_{C}$ remains connected, because, otherwise, we shall have
\begin{align*}
I(\mcH_{C})\utag{a}\leq 1\utag{b}=I(\mcH[\mcP^*(\mcH)])\utag{c}= I(\mcH)
\end{align*}
which contradicts~Proposition~\ref{pro:P*}. Here, \uref{a} can be argued in a similar manner as~\eqref{eq:I(H)<=1}; \uref{b} follows from~\eqref{eq:HT:I=1} since $\mcH[\mcP^*(\mcH)]$ is connected and cycle-free; \uref{c} follows from~\eqref{eq:I(H)} and Definition~\ref{def:H[P]} of $\mcH[\mcP^*(\mcH)]$. 
\item Connectedness outside $C$: the vertices in $\mcP^*(\mcH)$ remain connected either, since $\xi_{\mcH}(\rm{e})\subseteq \xi_{\mcH[\mcP^*(\mcH)]}(\rm{e})=C$ by Definition~\ref{def:H[P]} and $\mcH[\mcP^*(\mcH)]$ is a hypertree. 
\end{itemize}
Therefore, $\mcH$ remains connected after removing $\rm{e}$, contradicting the fact that $\mcH$ is minimally connected. Hence, $\mcH[\mcP^*(\mcH)]$ is a hypertree. 

Next, we proceed to prove~\eqref{eq:mch:C(H)=d(C)}. For any $C\in\mcP^*(\mcH)$, consider two distinct $C', C'' \in \mcP^*(\mcH)\setminus C$. If there is a path in $\mcH/C$ between any $v' \in C'$ and any $v'' \in C''$, then there is a path in $\mcH[\mcP^*(\mcH)]/C$ between $C'$ and $C''$ by the Definition~\ref{def:H[P]} of $\mcH[\mcP^*(\mcH)]$. The contrapositive statement implies 
\begin{align*}
\kappa(\mcH/C)\geq \kappa(\mcH[\mcP^*(\mcH)]/C).
\end{align*}
Note that $\mcH_{C'}$ is connected for all $C'\in \mcP^*(\mcH)$ as argued before. If  there is a path in $\mcH[\mcP^*(\mcH)]/C$ between $C'$ and $C''$, then there is a path in $\mcH/C$ between all $v'\in C'$ and all $v''\in C''$.
The contrapositive statement implies 
\begin{align*}
\kappa(\mcH/C)\leq \kappa(\mcH[\mcP^*(\mcH)]/C).
\end{align*}
Thus, we have 
\begin{align}
\label{eq:c(H/C)=c(H[P]/C)}
\kappa(\mcH/C)=\kappa(\mcH[\mcP^*(\mcH)]/C).
%		      &=\mathtt{d}_{\mcH[\mcP^*(\mcH)]}(C)
\end{align}
Since $\mcH[\mcP^*(\mcH)]$ is a hypertree, each incident edge of $C$ in $\mcH[\mcP^*(\mcH)]$ connects $C$ to a different connected component of $\mcH[\mcP^*(\mcH)]/C$, namely, 
\begin{align}
\label{eq:c(H[P]/C=deg_H[P](C)}
\kappa(\mcH[\mcP^*(\mcH)]/C)=\mathtt{d}_{\mcH[\mcP^*(\mcH)]}(C).
\end{align}
From Definition~\ref{def:H[P]} and~\eqref{eq:deg:c}, it follows that
\begin{align}
\label{eq:ded_H[P](C)=deg_H(C)}
\mathtt{d}_{\mcH[\mcP^*(\mcH)]}(C)&=\mathtt{d}_{\mcH}(C),\quad\forall C\in \mcP^*(\mcH).
\end{align}
Combining~\eqref{eq:c(H/C)=c(H[P]/C)},~\eqref{eq:c(H[P]/C=deg_H[P](C)} and~\eqref{eq:ded_H[P](C)=deg_H(C)} yields~\eqref{eq:mch:C(H)=d(C)} as desired.

It now remains  to prove~\eqref{eq:mch:d(C)=E}. Since $\mcH[\mcP^*(\mcH)]$ is a hypertree, by Lemma~\ref{lem:HT}, we have
%\begin{subequations}
\begin{align*}
I(\mcH[\mcP^*(\mcH)])&=1, \\
\mcP^*(\mcH[\mcP^*(\mcH)])&=\Set{\Set{C}\mid C\in \mcP^*(\mcH)}.
\end{align*}
%\end{subequations}
It then follows from~\eqref{eq:I(H)} that
\begin{align*}
1\kern-0.3em=\frac{1}{\abs{\mcP^*(\mcH)}-1}\kern-0.3em\left[\sum_{C\in \mcP^*(\mcH)}\kern-1em\mathtt{d}_{\mcH[\mcP^*(\mcH)]}(C)-\abs{E(\mcH[\mcP^*(\mcH)])}\kern-0.1em\right]
\end{align*}
Note that by Definition~\ref{def:H[P]}, we have
\begin{align}
\label{eq:|E(H[P])|=|E(H)|}
 \abs{E(\mcH[\mcP^*(\mcH)])}&=\abs{E(\mcH)}. 
\end{align}
Substituting~\eqref{eq:ded_H[P](C)=deg_H(C)} and~\eqref{eq:|E(H[P])|=|E(H)|} into the above equation yields the desired result~\eqref{eq:mch:d(C)=E}. Lemma~\ref{lem:MCH} is proved.
%\end{Proof}

\section{Alternative Converse Proof of Theorem~\ref{thm:MCH:Region}}
\label{sec:proof:converse:alternative}

To begin with, consider any $B\subseteq V$ with size $\abs{B}<|V|-1$. We want to identify the best $\mcP\in\Pi'(V\setminus B)$ that gives the tightest bound on $r(B)$ in~\eqref{eq:LB}. Consider the following two cases:

\underline{Case 1:} $\kappa(\mcH/B)=1$, i.e., $\mcH$ remains connected after removing all the vertices in $B$. For any $\mcP \in\Pi'(V\setminus B)$, we have
  \begin{align*}
	 I_{\mcP}(\RZ_{V\setminus B})\geq  \min_{\rm{e}\in E(\mcH/B)} w(\rm{e})\geq \min_{\rm{e}\in E(\mcH)} w(\rm{e}),
   \end{align*}
where the first inequality can be proved in the same manner as in the achievability proof of Proposition~\ref{pro:CS}; the second inequality is because $E(\mcH/B)\subseteq E(\mcH)$ by Definition~\ref{def:H/C}.
It then follows from~\eqref{eq:MCH:CS} that
\begin{align*}
 &\left(\abs{\mcP}-1\right)\left[\rK-I_{\mcP}(\RZ_{V\setminus B})\right]\\
 &\leq (\abs{\mcP}-1)\left[\CS(\infty)-I_{\mcP}(\RZ_{V\setminus B})\right]\\
 &\leq 0,\quad\forall\mcP \in\Pi'(V\setminus B).
 \end{align*}
This means~\eqref{eq:LB} is trivial for all $\mcP \in\Pi'(V\setminus B)$, and 
we use 
\begin{align*}
r(B)\geq \left[\kappa(\mcH/B)-1\right]\rK=0 
\end{align*}
instead.

\underline{Case 2:} $\kappa(\mcH/B)> 1$, i.e., $\mcH$ will become disconnected after removing the vertices in $B$.
%For the fundamental partition $\mcP^*(\mcH/B)$ of hypergraph $\mcH/B$, we have
%\begin{align}
%\label{eq:I_P*=0}
%|\mcP^*(\mcH/B)|=\kappa(\mcH/ B)\kern.5em \text{and}\kern.5em I_{\mcP^*(\mcH/B)}(\RZ_{V\setminus B})= 0
%\end{align}
Upon manipulating~\eqref{eq:LB}, we have
	\begin{align}
	\label{eq:psp}
		 r(B)&\geq (\abs{\mcP}-1)\left[\rK-I_{\mcP}(\RZ_{V\setminus B})\right] \notag\\
		       &=(\abs{\mcP}-1)\rK-\sum_{C\in\mcP}H(\RZ_C)+H(\RZ_{V\setminus B})\notag\\
		       &=-\sum_{C\in\mcP}\left[H(\RZ_C)-\rK\right]+H(\RZ_{V\setminus B})-\rK
	\end{align}
Now, consider $\mcP\in\Pi(V\setminus B)$ instead. For $C\in\mcP$ : $|C|>1$ and $\mcP^C\in\Pi'(C)$, we call $\mcP':=\mcP^C\cup\mcP\setminus \Set{C}$ is a \emph{refinement} of $\mcP$ through $C$. According to~\cite[Theorem~3.7]{narayanan90}, the optimal $\mcP$ that gives the largest value to the R.H.S. of~\eqref{eq:psp} lies in a special sequence 
\begin{align*}
\mcP_0\!=\!\Set{V\setminus B}, \mcP_1\!=\!\mcP^*(\mcH/B),\dots, \mcP_{i}\!=\!\Set{\Set{j}\mid j\in V\setminus B},
\end{align*}
where $\mcP_{\ell}$ is a refinement of $\mcP_{\ell-1}$ through some $C\in\mcP_{\ell-1}$, $1\leq\ell\leq i$. Next, we show that the optimal one is indeed the second one in the above sequence. For $\ell\geq 2$, it was shown by~\cite[Corollary~5.3]{chan15mi} that 
%\begin{subequations}
%\label{eq:Ip:chain}
\begin{align*} 
I_{\mcP_{\ell}}(\RZ_{V\setminus B})&=\theta I_{\mcP_{\ell-1}}(\RZ_{V\setminus B})+(1-\theta)I_{\mcP^{C}}(\RZ_C) 
\end{align*}
where $C\in\mcP_{\ell-1}, \mcP^{C}\in\Pi'(C)$, and
\begin{align*}
 \theta&=\frac{|\mcP_{\ell-1}|-1}{|\mcP_{\ell}|-1}=\frac{|\mcP_{\ell-1}|-1}{|\mcP_{\ell-1}|+|\mcP^C|-2}\in (0, 1).
\end{align*}
%\end{subequations}
Now, it follows that
\begin{align*}
&(\abs{\mcP_{\ell}}-1)\left[\rK- I_{\mcP_{\ell}}(\RZ_{V\setminus B})\right]\\
&\quad-(\abs{\mcP_{\ell-1}}-1)\left[\rK- I_{\mcP_{\ell-1}}(\RZ_{V\setminus B})\right]\\
&= (|\mcP^C|-1)[\rK-I_{\mcP^C}(\RZ_C)]\\
&\leq 0  
\end{align*}
for all $\ell\geq 2$, where the inequality can be argued as follows:
\indent Let $C'\in \mcP_1\setminus \mcP_2$. For any $ C\in \mcP_{\ell-1}: \ell \geq 2$, we have  
\begin{align*}
I_{\mcP^C}(\RZ_C)\utag{a}\geq I(\RZ_{C'}) &\utag{b}\geq \min_{\rm{e}\in E(\mcH_{C'})} w(\rm{e})\\ &\utag{c}\geq \min_{\rm{e}\in E(\mcH)} w(\rm{e})\utag{d}=\CS(\infty)\utag{e}\geq \rK
\end{align*}
\indent Here,
\begin{itemize}
\item \uref{a} follows from~\cite[Theorem~7]{chan16cluster}; 
\item \uref{b} is because $C'\in \mcP_1=\mcP^*(\mcH/B)$, by Proposition~\ref{pro:P*}, $I(\mcH_{C'})>I(\mcH/B)=0$, and so the corresponding hypergraph $\mcH_{C'}$ of $\RZ_{C'}$ is connected, and then we can apply the same method as in the achievability proof of Proposition~\ref{pro:CS} to derive the desired result;
\item \uref{c} follows from $E(\mcH_{C'})\subseteq E(\mcH)$ by Definition~\ref{def:H/C};
\item \uref{d} follows from~\eqref{eq:MCH:CS};
\item \uref{e} follows from~\eqref{eq:cs8}.
\end{itemize}
Therefore, $\mcP_1=\mcP^*(\mcH/B)$ is better than all $\mcP_\ell,\ell \geq2$. It remains to compare it with $\mcP_0=\Set{V\setminus B}$. For $\mcP_1$, we have
\begin{align*}
 &-\sum_{C\in\mcP_1}[H(\RZ_C)-\rK]+H(\RZ_{V\setminus B})-\rK\\
 &=(\abs{\mcP^*(\mcH/B)}-1)[\rK-I_{\mcP^*(\mcH/B)}(\RZ_{V\setminus B})]\\
 &=[\kappa(\mcH/B)-1]\rK\\
 &> 0
\end{align*}
where the last equality follows from
\begin{align*}
|\mcP^*(\mcH/B)|=\kappa(\mcH/ B)\kern.5em \text{and}\kern.5em I_{\mcP^*(\mcH/B)}(\RZ_{V\setminus B})= 0
\end{align*}
because, every hyperedge of the corresponding hypergraph $\mcH/ B$ of $\RZ_{V\setminus B}$ is entirely contained by a part of $\mcP^*(\mcH/B)$, i.e., no edges cross $\mcP^*(\mcH/B)$. For $\mcP_0$, we have
\begin{align*}
-\sum_{C\in\mcP_0}[H(\RZ_C)-\rK]+H(\RZ_{V\setminus B})-\rK=0
\end{align*}
Therefore, $\mcP_1=\mcP^*(\mcH/B)$ is the best one that gives the largest value to the R.H.S. of~\eqref{eq:psp}, thereby the tightest bound on $r(B)$.
%Therefore, for each $B$, we only need
%\begin{align*}
% r(B)&\geq (\abs{\mcP^*(\mcH/B)}-1)[\rK-I_{\mcP^*(\mcH/B)}(\RZ_{V`/B})]\\
%      &=[\kappa(\mcH/B)-1]\rK
%\end{align*}

Summarizing the above two cases, we have 
\begin{align}
r(B)\geq [\kappa(\mcH/B)-1]\rK, \quad\forall B\subsetneq V.\label{eq:r(B):V}
\end{align}
Then, we shall use the following technical Lemma~\ref{lem:subadditive} to identify the redundant inequalities in~\eqref{eq:r(B):V}.
\begin{Lemma}
For any $B\subseteq V(\mcH)$ of a MCH $\mcH$, we have
\label{lem:subadditive}
\begin{align}
\kappa(\mcH/B) \leq \sum_{i=1}^{q} \kappa(\mcH/B_i)-(q-1), %\forall B\subseteq V,
\end{align}
where $\mcP^*(\mcH)=\Set{C_1,\dots,C_q}$ with $q=|\mcP^*(\mcH)|$ is the fundamental partition of $\mcH$, and $B_i=B\cap C_i$ for $i=1,\dots,q$. %i.e., subadditive in $B\not\subseteq C\in \mcP^*(\mcH)$.
\end{Lemma}
\begin{Proof}
See Appendix~\ref{sec:proof:lem:subadditive}.
\end{Proof}
Lemma~\ref{lem:subadditive} asserts that $B\to\kappa(\mcH/B)-1$ is subadditive. This property is illustrated by the following simple example.  
\begin{Example} Let us consider the MCH $\mcH$ in~\figref{fig:mch}. Recall that $\mcP^*(\mcH)=\Set{\Set{1,2,3},\Set{4},\Set{5},\Set{6}}$. For $B=\Set{2,3,4}$,  we have  $B_1=\Set{2,3},B_2=\Set{4},B_3=\emptyset,B_4=\emptyset$. It is readily seen that $\kappa(\mcH/B)=\kappa(\mcH/B_1)=2$, $\kappa(\mcH/B_2)=\kappa(\mcH/B_3)=\kappa(\mcH/B_4)=1$. Therefore, 
\begin{align*}
\kappa(\mcH/B)-1 \leq \sum_{i=1}^{4} [\kappa(\mcH/B_i)-1] 
\end{align*} 
holds with equality.
%as implied by the above lemma.
Now, consider $B=\Set{1,2,4}$ instead. We have $B_1=\Set{1,2},B_2=\Set{4},B_3=\emptyset,B_4=\emptyset$. It is easy to see that $\kappa(\mcH/B)=\kappa(\mcH/B_2)=\kappa(\mcH/B_3)=\kappa(\mcH/B_4)=1$, $\kappa(\mcH/B_1)=2$. 
Hence, 
\begin{align*}
\kappa(\mcH/B)-1 \leq \sum_{i=1}^{4} \left[\kappa(\mcH/B_i)-1\right]
\end{align*} 
holds with strict inequality.
\end{Example}

Now, let us resume the converse proof. To invoke Lemma~\ref{lem:subadditive}, set $$q=|\mcP^*(\mcH)|  \kern1em\text{and}\kern1em \mcP^*(\mcH)=\Set{C_1,\dots,C_q}.$$ Consider any $B\subseteq V$ that satisfies $B\not\subseteq C_i, \forall i=1,\dots,q$. Define 
\begin{align*}
B_i=B\cap C_i,  \quad i=1,\dots,q.
\end{align*} 
Then, by Lemma~\ref{lem:subadditive}, we have
\begin{align*}
[\kappa(\mcH/ B)-1]\rK\leq \sum_{i=1}^{q}[\kappa(\mcH/ B_i)-1]\rK
\end{align*}
which implies the inequality in~\eqref{eq:r(B):V} that corresponds to $B$ is redundant. Therefore, we only need to consider the inequalities involving $B\subseteq C\in\mcP^*(\mcH)$ in~\eqref{eq:r(B):V}. This, in conjunction with
 the fact that $\rK\leq\CS(\infty)$, completes the converse proof of Theorem~\ref{thm:MCH:Region}. 
%\end{Proof}

\section{Proof of Lemma~\ref{lem:subadditive}}
\label{sec:proof:lem:subadditive}
%\begin{Proof}[Lemma~\ref{lem:subadditive}]
Consider a MCH $\mcH$, let $q=|\mcP^*(\mcH)|>1$. 
By the assertion~\ref{mch-to-hypertree} of Lemma~\ref{lem:MCH}, we know that $\mcH[\mcP^*(\mcH)]$ is a hypertree. 
Therefore, we can use the reordering method as in the proof of Lemma~\ref{lem:HT} to enumerate $\mcP^*(\mcH)$ as $\{C_1,\dots,C_q\}$ such that $C_{i+1}$ and $\bigcup_{j=1}^i C_j$ share exactly one edge for all $1\leq i\leq q-1$, i.e., satisfying~\eqref{eq:C:e:1}.
For any $B\subseteq V$, define $B_i:=B\cap C_i$ for $i=1,\dots,q$. Since $\mcH$ is connected, it follows that $\kappa(\mcH/B_i)=1$ for $B_i=\emptyset$. Now, suppose that $B$ intersects with $\ell$ parts of $\mcP^*(\mcH)$, where $2\leq\ell\leq q$ since the claim holds trivially when $\ell=1$. Define $T:=\Set{t\in[q]\mid B_t\neq\emptyset}$. Note that $|T|=\ell$. The claim is proved if we can show that
\begin{align}
	\label{eq:sub:nonempty}
\kappa(\mcH/B) \leq \sum_{t\in T} \kappa(\mcH/B_t)-(\ell-1).
\end{align}
We prove~\eqref{eq:sub:nonempty} by induction on $\ell$ for $\ell\geq 2$. First, assume $\ell=2$ and $T=\Set{t_1,t_2}$, i.e., $B\subseteq (C_{t_1}\cup C_{t_2})$ and $B\cap C_{t_1}\neq\emptyset, B\cap C_{t_2}\neq\emptyset$. Among those $\kappa(\mcH/B)$ connected components of $\mcH/B$, consider the following two cases: 

\underline{Case 1:} There is no connected component among those $\kappa(\mcH/B)$ such that it shares edge with both $B_{t_1}$ and $B_{t_2}$ in $\mcH$. Note that $\mcH$ is connected. It then follows that 
 \begin{align*}
\kappa(\mcH/B)&=\kappa(\mcH/(B_{t_1}\cup B_{t_2}))\\&= \kappa(\mcH/B_{t_1})-1+\kappa(\mcH/B_{t_2})-1\notag\\
 &\leq \kappa(\mcH/B_{t_1})+\kappa(\mcH/B_{t_2})-1.
 \end{align*}
 
 \underline{Case 2:} There exists connected component among those $\kappa(\mcH/B)$ such that it shares edge with both $B_{t_1}$ and $B_{t_2}$ in $\mcH$. Since $\mcH[\mcP^*(\mcH)]$ is a hypertree, there is only one such connected component. Upon noting $\mcH$ is connected, we have
 \begin{align*}
 \kappa(\mcH/B)&=\kappa(\mcH/(B_{t_1}\cup B_{t_2}))\\
 &= \kappa(\mcH/B_{t_1})+\kappa(\mcH/B_{t_2})-1.
 \end{align*}
Putting it all together, \eqref{eq:sub:nonempty} is ture for $\ell=2$.

Next, assume~\eqref{eq:sub:nonempty} is true for $\ell-1$, 
i.e., for some $T=\Set{t_1\dots,t_{\ell-1}}$,
\begin{align}
\label{eq:sub:induc_hypo}
\kappa\Bigg(\mcH/\bigcup_{j=1}^{\ell-1}B_{t_j}\Bigg) \leq \sum_{j=1}^{\ell-1} \kappa(\mcH/B_{t_j})-(\ell-2).
\end{align}
Then, consider the case that $B$ intersects with $\ell$ parts of $\mcP^*(\mcH)$. Since $C_{t_{\ell}}$ has exactly one edge with $\bigcup_{j=1}^{t_{\ell}-1}C_{j}$ and $\bigcup_{j=1}^{\ell-1}C_{t_j}\subseteq\bigcup_{j=1}^{t_{\ell}-1}C_{j}$, 
we have that $B_{t_\ell}$ and $\bigcup_{j=1}^{\ell-1}B_{t_j}$ can at most share one edge.

\underline{Case 1:}  $B_{t_\ell}$ and $\bigcup_{j=1}^{\ell-1}B_{t_j}$ share an edge $\rm{e}\in E(\mcH)$. After removing $\Set{B_{t_1},\dots, B_{t_\ell}}$, depending on whether $\xi_{\mcH}(\rm{e})\setminus \Set{B_{t_1},\dots, B_{t_\ell}}$ is empty or not, there can be at most one connected component that shares edge with both $B_{t_\ell}$ and $\bigcup_{j=1}^{\ell-1}B_{t_j}$ in $\mcH$. 

\underline{Case 2:}  $B_{t_\ell}$ and $\bigcup_{j=1}^{\ell-1}B_{t_j}$ do not share an edge. Since $\mcH[\mcP^*(\mcH)]$ is a hypertree, after removing $\Set{B_{t_1},\dots, B_{t_\ell}}$, there is exactly one connected component that shares edge with both $B_{t_\ell}$ and $\bigcup_{j=1}^{\ell-1}B_{t_j}$ in $\mcH$. 

Then, it follows that 
 \begin{align*}
 \kappa(\mcH/B)&=\kappa\Bigg(\mcH/\bigcup_{j=1}^{\ell}B_{t_j}\Bigg)\\
 &\leq \kappa\Bigg(\mcH/\bigcup_{j=1}^{\ell-1}B_{t_j}\Bigg)+\kappa(\mcH/B_{t_\ell})-1\\
 						   &\leq \sum_{j=1}^{\ell} \kappa(\mcH/B_{t_j})-(\ell-1)
 \end{align*}
 where the first inequality can be argued in a similar manner as the case $\ell=2$; the last inequality follows from the induction hypothesis~\eqref{eq:sub:induc_hypo}. This proves Lemma~\ref{lem:subadditive}.
%\end{Proof}

\section{Proof of Lemma~\ref{lem:H_Ec}}
\label{sec:proof:lem:H_Ec}
%\begin{Proof}[Lemma~\ref{lem:H_Ec}]
To begin with, consider any $C\in\mcP^*(\mcH)$ with size $|C|>1$. 
%We have $C \subsetneq V_C$, because. otherwise, $\mcH$ will become disconnected. 
By Propostion~\ref{pro:P*}, we have that
\begin{align*}
I(\mcH_C)>I(\mcH) \geq 0.
\end{align*}
This implies that $\mcH_C$ is connected, and so is $\mcH_{E_C}$ by its definition~\eqref{def:eq:H_Ec}. Furthermore, $\mcH_{E_C}$ is minimally connected, because, otherwise, it contradicts $\mcH$ is a MCH. As such, it is clear that
\begin{align}
\label{eq:H_Ec:ia}
\mathtt{d}_{\mcH_{E_C}}(i) \geq 1, \quad\forall i\in V_C.
\end{align}
Then, by the definition of $\mcH_{E_C}$ in~\eqref{def:eq:H_Ec} and~\eqref{eq:mch:C(H)=d(C)} in Lemma~\ref{lem:MCH}, we have 
\begin{align*}
\kappa(\mcH_{E_C}/C)\geq \kappa(\mcH/C)=\mathtt{d}_{\mcH}(C)=\mathtt{d}_{\mcH_{E_C}}(C),
\end{align*}
which, together with the fact that $\mcH_{E_C}$ is connected, yields
\begin{align}
\label{eq:H_Ec:ib}
\mathtt{d}_{\mcH_{E_C}}(i) =1, \quad\forall i\in V_C\setminus C.
\end{align}
Next, suppose to the contrary that there exists node in $C$ with degree one, i.e., $\exists i\in C$ such that $\mathtt{d}_{\mcH_{E_C}}(i)=\mathtt{d}_{\mcH}(i) =1$. It follows that 
\begin{align*}
\mathtt{d}_{\mcH}(C)=\mathtt{d}_{\mcH}(C\setminus\!\{i\})+\mathtt{d}_{\mcH}(i)-1
\end{align*}
since $|C|>1$ and $\mcH_C$ is connected.
On the other hand, by~\eqref{eq:mch:d(C)=E} in Lemma~\ref{lem:MCH}, it is clear that
\begin{align*}
I(\mcH)=\frac{\sum_{C\in\mcP^*(\mcH)}\mathtt{d}_{\mcH}(C)-|E(\mcH)|}{|\mcP^*(\mcH)|-1}=1.
\end{align*}
For notational simplicity, let $\mcP^*=\mcP^*(\mcH)$.  Define $$\mcP=\Set{C\setminus\! \{i\},\Set{i}}\cup \mcP^*\setminus \{C\}.$$ It follows that $\mcP\preceq \mcP^*$ and
\begin{align*}
&\frac{1}{|\mcP|-1}E_{\mcP}(\mcH)\\
&=\frac{\sum_{C'\in \mcP^*\setminus \{C\}}\mathtt{d}_{\mcH}(C')+\mathtt{d}_{\mcH}(C\setminus\! \{i\})+\mathtt{d}_{\mcH}(i)-|E(\mcH)|}{|\mcP^*|+1-1}\\
&=\frac{\sum_{C'\in \mcP^*}\mathtt{d}_{\mcH}(C')-|E(\mcH)|+1}{|\mcP^*|}\\
&=1\\
&=I(\mcH)
\end{align*}
which contradicts $\mcP^*$ is the fundamental partition. Hence,
\begin{align}
\label{eq:H_Ec:ic}
\mathtt{d}_{\mcH_{E_C}}(i) > 1, \quad\forall i\in C.
\end{align}
Upon combining~\eqref{eq:H_Ec:ia},~\eqref{eq:H_Ec:ib} and~\eqref{eq:H_Ec:ic}, we obtain the assertion~\ref{H_Ec:i} as desired. 

Finally, it remains to prove the assertion~\ref{H_Ec:ii}. Upon noting that $\mcH[\mcP^*(\mcH)]$ is a hypertree from the first assertion of Lemma~\ref{lem:MCH}, and therefore loopless, we have
\begin{align*}
\xi_{\mcH}(\rm{e})\setminus C\neq\emptyset, \quad\forall \rm{e}\in E_C.
\end{align*} 
Then, by the assertion~\ref{H_Ec:i} argued above, we get 
\begin{align*}
\forall \rm{e}\in E_C,\,\exists i,j\in  \xi_{\mcH_{E_C}}(\rm{e}) \,\,\text{s.t.}\,\, \mathtt{d}_{\mcH_{E_C}}(i)=1<\mathtt{d}_{\mcH_{E_C}}(j),
\end{align*} 
thereby establishing the assertion~\ref{H_Ec:ii}. The proof is completed. %Lemma~\ref{lem:H_Ec}.
%\end{Proof}

\section{Proof of Lemma~\ref{lem:supermodular}}
\label{sec:proof:lem:supermodular}
%\begin{Proof}[Lemma~\ref{lem:supermodular}]
To begin with, consider any $C\in\mcP^*(\mcH)$. By the definition of $\mcH_{E_C}$ in~\eqref{def:eq:H_Ec}, we have 
\begin{align*}
\kappa(\mcH_{E_C}/ B) \geq \kappa(\mcH/ B), \quad\forall B\subseteq C.
\end{align*}
On the other hand, the connected components of $\mcH_{E_C}/ B$ still remain disconnected in hypergraph $\mcH/ B$, i.e., 
\begin{align*}
\kappa(\mcH_{E_C}/ B) \leq \kappa(\mcH/ B), \quad\forall B\subseteq C,
\end{align*}
because, otherwise, upon noting $\mcH_{E_C}$ is connected by~Lemma~\ref{lem:H_Ec}, there exists some $C'\in\mcP^*(\mcH)\setminus\{C\}$ having two paths in $\mcH$ to $B$, thereby to $C$, however, this will contradict $\mcH[\mcP^*(\mcH)]$ is a hypertree by the first assertion of Lemma~\ref{lem:MCH}. 
Therefore, we get
\begin{align}
\label{eq:cc:H_Ec/B=H/B}
 \kappa(\mcH/ B) = \kappa(\mcH_{E_C}/ B), \quad\forall B\subseteq C.
\end{align}
As such, it suffices to show that $\kappa(\mcH_{E_C}/ B)$ is supermodular. Without loss of generality, let 
\begin{align*}
\mathcal{I}_{C}=\left\{1,\dots,\kappa(\mcH_{E_C}/ C)\right\}\subsetneq V_C
\end{align*}
be the set of representatives of the connected components in hypergraph $\mcH_{E_C}/C$. It follows from Lemma~\ref{lem:H_Ec}~\ref{H_Ec:i} that 
\begin{align*}
\mathtt{d}_{\mcH_{E_C}}(i)=1, \quad\forall i\in \mathcal{I}_{C}.
\end{align*} 
It also suffices to only consider the case $V_C=\mathcal{I}_{C}\cup C$, because
\begin{align*}
\kappa(\mcH_{E_C}/B)=\kappa(\mcH_{E_C}/[V_C\setminus (\mathcal{I}_{C}\cup C)\cup B]), \quad\forall B\subseteq C.
\end{align*} 
For $B\subseteq C$ and $i, j\in \mathcal{I}_{C}$, we write $i\sim_{\mcH_{E_C}/B} j$ to indicate that $j$ is reachable from $i$ via a path in hypergraph $\mcH_{E_C}/B$. Note that $\sim_{\mcH_{E_C}/B}$ is an equivalence relation and we denote the set of equivalence classes as
\begin{align*}
\mcP_B(\mathcal{I}_{C})=\op{maximal}\left\{\left.S\subseteq \mathcal{I}_{C}\right| i\sim_{\mcH_{E_C}/B} j, \forall i,j\in S\right\}.
\end{align*}
Now, it follows that
\begin{align}
\label{eq:cc:classe}
\kappa(\mcH_{E_C}/B)=|\mcP_B(\mathcal{I}_{C})|
\end{align}
because, by~\ref{H_Ec:ii} of Lemma~\ref{lem:H_Ec}, every edge must contain a vertex of degree one, which, by~\ref{H_Ec:i} of Lemma~\ref{lem:H_Ec}, is a representative in $\mathcal{I}_{C}$. Hence, it suffices to show that the R.H.S. of~\eqref{eq:cc:classe} is supermodular. Towards this end, consider any $S,T\subseteq C$. First, observe that for $i, j\in \mathcal{I}_{C}$, 
\begin{align*}
i\not\sim_{\mcH_{E_C}/S} j \kern.5em\text{or} \kern.5em i\not\sim_{\mcH_{E_C}/T} j \Longrightarrow i\not\sim_{\mcH_{E_C}/(S\cup T)} j.
\end{align*}
We then consider the following two cases:

\underline{Case 1:} $\forall i,j\in \mathcal{I}_{C}$ satisfying $i\not\sim_{\mcH_{E_C}/S} j$ and $i\not\sim_{\mcH_{E_C}/T} j$, we have $i\not\sim_{\mcH_{E_C}/(S\cap T)} j$. For this case, we shall have $S\cap T\neq \emptyset$, because, otherwise, $\mcH_{E_C}$ will become disconnected, which contradicts Lemma~\ref{lem:H_Ec}. Now, it follows that
  \begin{align}
  \label{eq:supermodular:=}
  |\mcP_S(\mathcal{I}_{C})|+|\mcP_T(\mathcal{I}_{C})|= |\mcP_{S\cup T}(\mathcal{I}_{C})|+|\mcP_{S\cap T}(\mathcal{I}_{C})|.
  \end{align}
  
\underline{Case 2:} $\exists i,j\in \mathcal{I}_{C}$ with $i\not\sim_{\mcH_{E_C}/S} j$ and $i\not\sim_{\mcH_{E_C}/T} j$, but $i\sim_{\mcH_{E_C}/(S\cap T)} j$. Then, by~\ref{H_Ec:ii} of Lemma~\ref{lem:H_Ec}, there exists at least a $u\in \mathcal{I}_{C}$ such that $i\not\sim_{\mcH_{E_C}/(S\cup T)} u\not\sim_{\mcH_{E_C}/(S\cup T)} j$, but $i\sim_{\mcH_{E_C}/T} u$ and $j\sim_{\mcH_{E_C}/S} u$. Now, we have
  \begin{align}
  \label{eq:supermodular:<=}
  |\mcP_S(\mathcal{I}_{C})|+|\mcP_T(\mathcal{I}_{C})|\leq |\mcP_{S\cup T}(\mathcal{I}_{C})|+|\mcP_{S\cap T}(\mathcal{I}_{C})|.
  \end{align}

Upon combining~\eqref{eq:supermodular:=} and~\eqref{eq:supermodular:<=}, we have that the R.H.S. of~\eqref{eq:cc:classe} is supermodular, and so is $\kappa(\mcH/ B), B\subseteq C\in\mcP^*(\mcH)$, by~\eqref{eq:cc:H_Ec/B=H/B}. This completes the proof of Lemma~\ref{lem:supermodular}.
%\end{Proof}

\section{Proof of Corollary~\ref{lem:extreme_point}}
\label{sec:proof:lem:extreme_point}

%\begin{Proof}[Corollary~\ref{lem:extreme_point}]
By virtue of Edmond's theorem~\cite[Corollory 44.3e]{schrijver02} concerning the extreme points of contra-polymatroid, every extreme point is expressed as
\begin{align*}
r_{\pi(i_1)}&=f(\Set{\pi(i_1)})=[\kappa(\mcH/\Set{\pi(i_1)})-1]\rK,\\
r_{\pi(i_\ell)}&=f(\Set{\pi(i_1),\dots,\pi(i_\ell)})-f(\Set{\pi(i_1),\dots,\pi(i_{\ell-1})})\\
	    %&=[\kappa(\mcH/\Set{\pi(i_1),\dots,\pi(i_\ell)})-1]\rK \\ 
	    %&\kern3em-[\kappa(\mcH/\Set{\pi(i_1),\dots,\pi(i_{\ell-1})})-1]\rK \\
	    &=[\kappa(\mcH/\Set{\pi(i_1),\dots,\pi(i_\ell)})\\
	    &\kern3em-\kappa(\mcH/\Set{\pi(i_1),\dots,\pi(i_{\ell-1})})]\rK
\end{align*}
for $\ell=2,\dots, |C|$, where $\pi(i_1),\pi(i_2),\dots,\pi(i_{|C|})$ is a permutation of $i_1,i_2,\dots,i_{|C|}$. 
This completes the assertion~\ref{pp:Pc:extreme_point}. The assertion~\ref{pp:Pc:internal_point} follows immediately from a general property of contra-polymatroid~\cite{Edmonds}. This proves Corollary~\ref{lem:extreme_point}.
\end{document}